# Realization Utility

# with Reference-Dependent Preferences[†]

**Jonathan E. Ingersoll, Jr.**

Yale School of Management

**Lawrence J. Jin**

Yale School of Management

PO Box 208200
New Haven CT
06520-8200

203-432-5924

Jonathan.Ingersoll@Yale.edu
Jin@Alumni.Caltech.edu

September 18, 2012

Forthcoming, *The Review of Financial Studies*

[†]We thank our colleagues at the Yale School of Management and the seminar participants at the LBS Trans-Atlantic Doctoral Conference, the University of Massachusetts, Amherst, and the 2012 Western Finance Association Annual Meeting for helpful discussions. In particular we would like to thank Nick Barberis, David Hirshleifer, and an anonymous referee. Jin acknowledges support from a Whitebox Advisors grant.

# Realization Utility with Reference-Dependent Preferences

## Jonathan E. Ingersoll, Jr. and Lawrence J. Jin

Yale School of Management

We develop a tractable model of realization utility that studies the role of reference-dependent S-shaped preferences in a dynamic investment setting with reinvestment. Our model generates both voluntarily realized gains and losses. It makes specific predictions about the volume of gains and losses, the holding periods, and the sizes of both realized and paper gains and losses that can be calibrated to a variety of statistics, including Odean's measure of the disposition effect. Our model also predicts several anomalies including, among others, the flattening of the capital market line and a negative price for idiosyncratic risk.

# Realization Utility with Reference-Dependent Preferences

## Jonathan E. Ingersoll, Jr. and Lawrence J. Jin

How do investors decide about and evaluate their own investment performance? Standard economic theory posits that investors maximize the expected utility of their lifetime consumption stream by dynamically adjusting their portfolio allocations based on their current wealth and their expectations of the future. Although this way of modeling investors' behavior may be close to reality for sophisticated investors, it is questionable whether less sophisticated investors behave this way.

A growing amount of research shows that individual investors do not always behave in the ways that expected utility theory predicts. In particular, the independence axiom seems troublesome as does the assumption of risk aversion, at least for losses. The latter assumption is not a requirement of expected utility theory, but it or something like it is important for most equilibrium models which follow from maximizing behavior. In contrast to what theories like the APT or CAPM predict, individual investors seem to be particularly concerned about an asset's change in price from a reference point.

Behavioral literature suggests alternative views of modeling investors' behavior. Kahneman and Tversky (1979) have argued that risk aversion does not characterize many choices and proposed an S-shaped utility function. Shefrin and Statman (1985) use "mental accounting" to justify investors' concentrating on specific separate incidents. Thaler (1999) says "A realized loss is more painful than a paper loss." Barberis and Xiong (2012) study a model which assumes that investors think of their investing experience as a series of separate episodes during each of which they either made or lost money and that the primary source of utility comes in a burst when a gain or loss is realized.[1] Frydman et al. (2012) find evidence using the neural data that supports this "realization utility" hypothesis.

In this paper, we use these notions to develop an intertemporal model of investors who have prospect theory's S-shaped utility and who evaluate their performance incident by incident based on realized profits and losses.[2] Our model is a partial equilibrium framework with an infinite horizon. An investor purchases stocks whose prices evolve as geometric Brownian motions. At each subsequent point in time, the investor decides whether to hold onto his current investment or realize his gain or loss thereby obtaining an immediate utility burst. If he sells, he reinvests the proceeds after transaction costs into another stock. We show that the investor's optimal strategy is to wait until the stock price rises or falls to certain percentages above or below the purchase price before selling. Our model includes that of Barberis and Xiong (2012) as a special case.

Voluntary loss taking can be optimal in a dynamic setting because the subsequent reinvestment resets the reference level and increases the likelihood of realizing future gains. But in the Barberis Xiong (2012) model, utility is piecewise linear. As a direct consequence, they predict that investors voluntarily realize gains but never voluntarily sell at a loss, which is clearly unrealistic and inconsistent with the data. In our model, with an S-shaped function, marginal utility decreases with the magnitude of both gains and losses. This means that lifetime utility can be increased by taking frequent small gains along with occasional larger losses because the latter have less total disutility than the utility of the former, and realizing losses resets the reference level for future gains. The disposition effect, an empirically robust pattern that individual investors have higher propensities to realize gains than to realize losses,[3] follows naturally from this result but it is a *dynamic* result.

---

[1] Further discussion on the psychological foundation of viewing investments as episodes is in their paper.

[2] Kyle, Ou-Yang, and Xiong (2006) and Henderson (2012) study one-time liquidation problems with prospect theory preferences. But reinvestment, which is a key component of our model, is ignored in their models.

What has been commonly argued in both theoretical and empirical literatures is that an S-shaped utility function leads to the disposition effect because risk seeking over losses induces investors to retain their positions and gamble on the future while risk aversion over gains induces the opposite. However, this is a *static* argument. Extrapolating this reasoning period by period would imply losses are never realized; the disposition effect should be infinite. In our dynamic realization utility model exactly the opposite is true. Investors naturally want gains, but an S-shaped utility helps to generate voluntary losses and thereby reduces the magnitude of the disposition effect to the observed level.

We calibrate our model in two parts. First we show that the magnitudes and frequencies of realized gains and losses and the frequencies of paper gains and losses as observed in the trading data of Odean and others are consistent with the type of simple two-point strategy our model predicts. In particular, Odean reports that 54% of round-trip trades are realized gains with an average size of 28%; the remainder are losses averaging –23%. Also conditional on a trade, investors realize 15% of possible gains and only 10% of possible losses. Using those average realized gain and loss sizes, our model makes the very accurate prediction that 58% of sales should be gains, and investors should have propensities of 14% and 11% to realize gains and losses, respectively. In addition, we propose a modified form of Tversky-Kahneman utility that generates the two optimal sales points, 28% and –23%, either alone or in a mixture of heterogeneous investors.

Our model also has a variety of other empirical implications and predictions. For instance, investors may be risk-seeking in some circumstances due to the option value inherent in realizing losses; this helps explain a flatter security market line and the negative pricing of idiosyncratic risk as shown in Ang, Hodrick, Xing, and Zhang (2006). It may also help explain why investors appear to hold portfolios that appear under-diversified.

The plan of our paper is as follows. In Section 1, we lay out a specific intertemporal reference-dependent realization utility model and present its solution and basic insights. Section

---

[3] The disposition effect for individual investors has been found in the U.S., Israel, Finland, China, and Sweden, by Odean (1998), Shapira and Venezia (2001), Grinblatt and Keloharju (2001), Feng and Seasholes (2005), and Calvet, Campbell, and Sodini (2009), respectively. It is also documented for U.S. mutual fund managers, the real estate market, and the exercise of executive stock options, by Frazzini (2006), Genesove and Mayer (2001), and Heath, Huddart, and Lang (1999).

2 examines the properties of the derived value function and analyzes the optimal sales policies. Section 3 provides a detailed calibration of our model to several empirical regularities. Section 4 analyzes voluntary loss realization in a more general context. Section 5 presents further model applications and predictions. Section 6 gives some concluding remarks and direction for future research. A summary of the important notation, all the proofs, and some more detailed technical considerations are included in the Appendix.

## 1. A Realization Utility Model with Tversky-Kahneman Utility

In this section we present a simple, specific model of intertemporal realization utility. Our investor takes positions in a series of purchases, buying a number of shares and later selling his entire position and reinvesting it.[4] Each realized gain or loss contributes a burst of utility, and our investor acts to maximize the expectation of the sum of the discounted values of these bursts.

We assume the investor applies narrow framing when he evaluates his gains. This assumption side-steps any complications that might arise from diversification or rebalancing motives.[5] Narrow framing can be justified if the investor derives realization utility only when both the purchase and sale prices of the asset are salient, and, therefore, evaluating *individual* assets is the applicable setting for studying realization utility. As a result, even when the investor holds multiple stock positions simultaneously, narrow framing allows us to study each sequence of purchases and sales separately.

Secondly, we assume that a utility burst is received only at the time when a gain or a loss is realized. As with prospect theory, we normalize utility so that gains and losses contribute positive and negative utility, respectively.[6] While it is assumed that utility depends primarily on the size, $G$ of the gain or loss, it seems reasonable that the reference level, $R$, might also have a separate effect. In particular, a gain or loss of a given size probably has a greater utility impact, either good or bad, the smaller is the reference level; e.g., the gain or loss of $10 is felt more strongly when the reference level is $100 than when it is $500. Therefore, we denote the utility burst function as a function of both variables, $U(G, R)$. In this paper we assume that $U(G, R)$ is homogeneous of degree β in $G$ and $R$

$$U(G,R) = R^{\beta} u(G/R). \tag{1}$$

---

[4] Our model restriction of full liquidation is an empirically plausible one for individual investors. Feng and Seasholes (2005) document that individual investors trading through a large Chinese brokerage house during 1999–2000 liquidated their full position 80.35% of the time when selling. Shapira and Venezia (2001) report that approximately 80% of round trips on the Tel Aviv Stock Exchange in 1994 consisted of a single purchase followed by a single sale of the entire holding. Kaustia (2010) reports a similar result for his Finnish data though he does not provide specific numbers.

[5] For instance, the investor might have an incentive to sell a losing stock to purchase a winning stock. Another example could be the incentive of purchasing a diversified fund. These considerations are outside the scope of this paper though some related discussion is provided in the last two sections of the paper.

[6] Typically the centering of utility is arbitrary and has no effect on expected utility maximization. In some models like this, the investor might be able to choose to take no action at all so if no action is presumed to give a utility of zero, then the centering chosen here can affect participation in the market.

This assumption is important for keeping the model tractable, but it also focuses utility on rates of return rather than dollar changes which is in keeping with the general emphasis in Finance. Expressed in this way, the scaling parameter β gauges the impact of the reference level on utility bursts measured as rates of return.

We study a Merton-type partial equilibrium economy in continuous time with an infinite horizon. At $t = 0$, the investor chooses either to stay out of the market which earns him a utility of zero or to invest in one of a number of identically distributed stocks. The stock price evolves according to a geometric Brownian motion, $dS/S = \mu dt + \sigma d\omega$, where μ and σ are the growth rate and logarithmic standard deviation, respectively, and ω is a standard Brownian motion. At each subsequent point in time, $t$, the investor chooses either to hold his investment for a longer time or to sell his entire position and realize a utility burst. When he sells, he pays a proportional transaction cost, $k_s$, and reinvests the net proceeds after paying a second proportional transaction cost, $k_p$ reducing his investment to $X_{t^+} = (1-k_s)X_{t^-}/(1+k_p) \equiv KX_{t^-}$. Between realization dates, the investment value follows the same geometric Brownian motion as the underlying asset

$$dX/X = \mu dt + \sigma d\omega. \qquad (2)$$

In a static prospect theory setting, the reference level is essentially a parameter of the utility function defining the status quo. However, in our dynamic model, we must address how it is updated and exactly how the gain or loss is measured relative to it. The simplest rule is $R$ is set at the net purchase price, as defined above, and remains constant between sales.[7] That is, when the investor sells his stock for $X_t$, he resets his reference level to $KX_t$ until the next sale. However, this is a subjective matter and could differ from investor to investor; there are other ways that the new reference level might be set. For instance, an investor might view it as the gross amount invested including the purchasing cost, i.e., $R = (1-k_s)X_t$. It might also be some intermediate level particularly if the transaction costs have different components such as a bid-ask spread and a commission. Most brokerage accounts show the purchase price which would tend to emphasize the net investment as the reference level. On the other hand, the tax cost basis includes the purchasing cost which would tend to emphasize the gross investment as the reference level. In our analysis, we assume the simplest case that investor fully accounts for costs and sets the reference level to the net amount invested, $KX$.[8]

A related issue is how the investor evaluates his utility burst upon a sale. Again, there are several ways he might do so depending on his subjective view of the transaction costs. For example, if he ignores costs completely, then the gain is the gross sales value less the reference level, $G_t = X_{t^-} - R_{t^-}$. If he fully recognizes transaction costs and compares the net reinvested amount to the reference level, then $G_t = KX_{t^-} - R_{t^-}$. If he views the gain as the difference between the net proceeds of the sale and the reference level, then $G_t = (1-k_s)X_{t^-} - R_{t^-}$. These three cases are covered by defining the gain as $G_t = \kappa X_{t^-} - R_{t^-}$, and setting the parameter $\kappa$ to 1, $K$, or $1-k_s$, respectively. Intermediate views are also possible. We leave the parameter $\kappa$ free allowing many interpretations.

---

[7] Throughout this paper we assume the reference level is constant between sales. More generally, it might grow deterministically at a constant rate (like the interest rate), or evolve stochastically over time. It could also be updated based on recent history of the stock price.

[8] The analysis here is largely unchanged for different ways of setting the reference level. If an investor adopts the gross cost view, then equation (5) below has $V(KX, (1-k_s)X)$ as the second term on the right-hand side. Presumably, an investor would not adopt a gross cost view for setting the new reference level as well as recognizing both costs in assessing the gain as this would double count the purchasing costs. However, the only requirement for our model is that the investor sets his reference level *consistently* over time. Barberis and Xiong (2012) also adopt the net cost interpretation which in their notation is $(1-k)X$ with $k$ being the round-trip transaction cost and only consider the full recognition of transaction costs in determining gain size.

The time-consistency of these rules together with the assumptions that (i) the utility bursts in (1) are homogeneous of degree $\beta$ in $X$ and $R$, (ii) the asset value process has stochastic constant returns to scale, and (iii) the investment horizon is infinite jointly guarantee that the future looks the same depending only on the current investment and reference level. This simplifies our problem in two ways. First, it removes time as an explicit variable. Second, our investor always has the incentive to reinvest immediately upon selling a position since he chose to enter the market in the first place.

Denote the value function discounted to time $t$ by $V(X_t, R_t)$. As discussed above, $V$ does not depend on time explicitly but only on the current investment and reference level. By definition, the value function is the maximized expectation of the sum of future discounted utility bursts

$$V(X_t, R_t) = \max_{\{\tilde{\tau}_i\}} \mathbb{E}_t \left[ \sum_i e^{-\delta \tilde{\tau}_i} U(\tilde{G}_{t+\tilde{\tau}_i}, \tilde{R}_{t+\tilde{\tau}_i}) \right] \tag{3}$$

where $\delta$ is the rate of time preference, $\tilde{G}_{t+\tilde{\tau}_i}$ and $\tilde{R}_{t+\tilde{\tau}_i}$ are the dollar size and the reference level for the $i^{th}$ future gain, respectively, and $t + \tilde{\tau}_i$ is the random time it is realized. In our model, these are stopping times that are endogenously chosen by the investor to maximize his lifetime expected utility.[9] To solve the problem posed by (3), we use the time-homogeneity property to rewrite it as a recursive expression

$$V(X_t, R_t) = \max_{\tilde{\tau}} \mathbb{E}_t \left[ e^{-\delta \tilde{\tau}} U(\kappa \tilde{X}_{t+\tilde{\tau}} - R_t, R_t) + e^{-\delta \tilde{\tau}} V(K \tilde{X}_{t+\tilde{\tau}}, K \tilde{X}_{t+\tilde{\tau}}) \right] \tag{4}$$

where $\tilde{\tau}$ is the time until the next sale. Hereafter, we suppress time subscripts for notational convenience unless necessary for clarity.

At a sale, the value function before the sale must equal the sum of the utility burst of the sale and the post-reinvestment continuation value function. So upon a sale realizing $X$ before costs,

$$V(X, R) = U(\kappa X - R, R) + V(KX, KX). \tag{5}$$

Between sales times, equation (4) can be re-expressed using the law of iterated expectations and Itô's lemma

$$0 = \mathbb{E}\{d[e^{-\delta t} V(X, R)]\} = e^{-\delta t} \left( \tfrac{1}{2} \sigma^2 X^2 V_{XX} + \mu X V_X - \delta V \right) dt. \tag{6}$$

---

[9] To complete the specification of this maximization, we need to assign a utility value to the *policy* of never executing any sales. The obvious choice in this case is to assign this policy a utility value of zero, the same value that would be realized with a policy that allowed sales but never happened to execute any.

Because $U(G, R)$ is homogeneous of degree $\beta$ in $G$ and $R$ and the asset value process has stochastic constant returns to scale, $V$ must be also homogeneous of degree $\beta$ in $X$ and $R$ and therefore can be written as $V(X, R) = R^\beta v(x)$, where $v$ is the reduced-form value function and $x \equiv X/R$ is the gross return per dollar of the reference value.[10] The equation for $v$ is

$$0 = \tfrac{1}{2}\sigma^2 x^2 v'' + \mu x v' - \delta v. \tag{7}$$

The general solution to (7) is

$$v(x) = C_1 x^{\gamma_1} + C_2 x^{\gamma_2} \quad \text{where} \quad \gamma_{1,2} \equiv \frac{-\mu + \tfrac{1}{2}\sigma^2 \pm \sqrt{(\mu - \tfrac{1}{2}\sigma^2)^2 + 2\delta\sigma^2}}{\sigma^2}. \tag{8}$$

This is true regardless of the form for $u$. The utility of the sales bursts affects only the constants, $C_1$ and $C_2$. Again due to the homogeneity, the optimal sales strategy must be to realize a gain or loss when the stock price reaches a constant multiple, $\Theta$, or fraction, $\theta$, of the reference level.[11] The upper sales point, $\Theta$, must exceed $1/\kappa > 1$ as otherwise the sale is not a gain after costs. The lower sales point, $\theta$, must be less than 1.[12]

Applying the homogeneity relation (1) to the boundary condition (5) yields the reduced-form boundary conditions

$$v(\Theta) = u(\kappa\Theta - 1) + (K\Theta)^\beta v(1), \qquad v(\theta) = u(\kappa\theta - 1) + (K\theta)^\beta v(1). \tag{9}$$

Equating these to the general solution from (8), we can determine the constants $C_1$ and $C_2$ in terms of the policy variables

$$\begin{aligned} C_1 &= \frac{c_2(\theta)u(\kappa\Theta - 1) - c_2(\Theta)u(\kappa\theta - 1)}{c_1(\Theta)c_2(\theta) - c_1(\theta)c_2(\Theta)} \\ C_2 &= \frac{c_1(\Theta)u(\kappa\theta - 1) - c_1(\theta)u(\kappa\Theta - 1)}{c_1(\Theta)c_2(\theta) - c_1(\theta)c_2(\Theta)} \quad \text{where} \quad c_i(\phi) \equiv \phi^{\gamma_i} - (K\phi)^\beta. \end{aligned} \tag{10}$$

The optimal sales points, $\Theta$ and $\theta$, can now be determined either by maximizing $C_1 + C_2$,

---

[10] $V$ must be positively homogeneous; i.e., $\beta \geq 0$. Since $V(X_0, X_0) = X_0^\beta v(1)$, utility is decreasing in the amount of the original investment when $\beta < 0$, and the investor would always prefer to reduce his initial investment and in the limit not participate at all. A positive $\beta$ also assures that $|U(gR, R)|$ is increasing in $R$ for a fixed $g$; that is, the higher the reference level, the bigger is the utility of a given rate of return. This property is similar to increasing relative risk aversion.

[11] See the Appendix for more details on the constancy of the optimal policy.

[12] A sale at any point in the range $(1, 1/\kappa)$ produces a subjective loss after accounting for transaction costs. Under a constant policy with $\theta$ in this range, there would never be any sales at a higher price as the stochastic process for $x$ is continuous and begins at 1 after each repurchase when the reference level is set to the net investment. But this means that only losses with their negative utility bursts would be realized leading to a negative $v$. This could not be the optimal policy as never selling gives a utility of zero as does not participating at all.

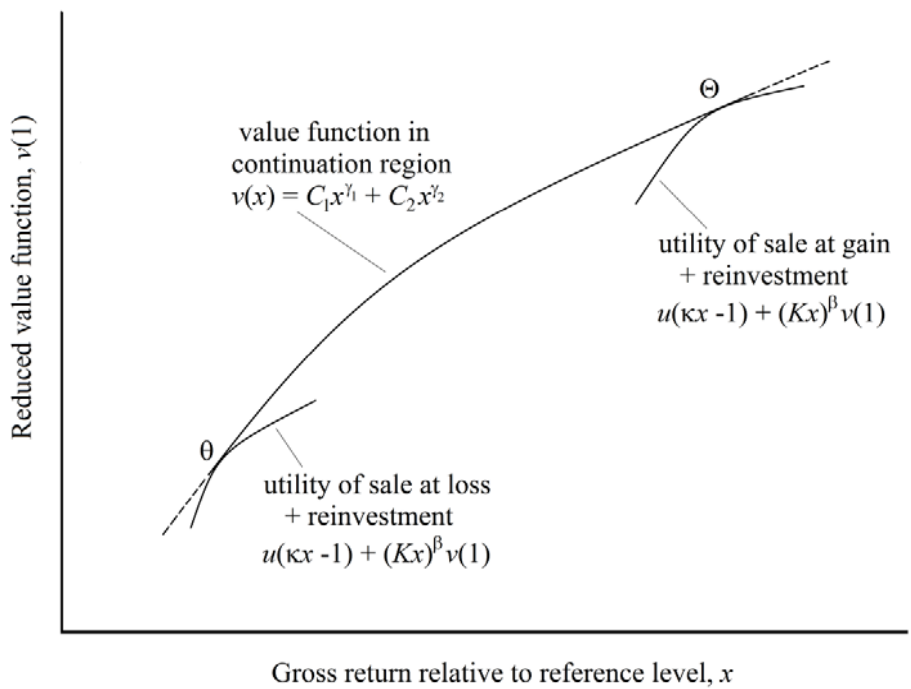

**Figure 1: Determination of the Optimal Sales Policies.** This figure illustrates the value function and the optimal policy for realizing gains and losses. The value in the continuation or no-sales region is tangent to the sum of the payoff function and the continuation value at each sales point, $\theta$ or $\Theta$.

---

which is the value of $v(1)$, or by applying the smooth-pasting condition at both sales points.[13] The solution and the optimal $\theta$-$\Theta$ strategy are illustrated in Figure 1. The no-sales region runs from $\theta$ to $\Theta$. The value function exceeds the sum of the utility burst plus the continuation value in this region as illustrated; it is tangent to the payoff including continuation value at $\Theta$ and $\theta$. For some parameter values, it may be optimal to forgo all losses. In these cases, the continuation value is not large enough to offset that disutility of realizing a loss. The value function is still given by (8) though $C_2 = 0$ and, therefore, $v(0) = 0$.

Typical stock and transaction costs values are $\mu = 9\%$, $\sigma = 30\%$, and $k_s = k_p = 1\%$. How would a realization-utility investor trade this stock? To answer this question, we must specify the burst utility function. A reference-scaled version of the Cumulative Prospect Theory (CPT)

---

[13] The optimal sales strategy must maximize $v$ for every value of its argument in the continuation region, and $x = 1$ is guaranteed to be in the continuation region since $\theta < 1 < 1/\kappa < \Theta$. Note that the smooth pasting condition does not simply match the derivative of $v$ to the marginal utility of the burst. It must be applied to (9) which has the continuation value as well as the utility burst on the right-hand side. As discussed in Proposition 1 below, in some cases there is a constrained optimum, $\theta = 0$, at which the smooth pasting condition does not apply. Unless $\delta > 0$ and $\beta \leq \gamma_1$, there is no unique optimum as many strategies lead to infinite utility. These transversality issues are discussed in the Appendix.

utility proposed by Tversky and Kahneman (1992), hereafter called scaled-TK utility, is

$$U_{sTK}(G,R) = R^{\beta} u_{sTK}(G/R) \quad \text{for} \quad u_{sTK}(g) = \begin{cases} g^{\alpha_G} & g \geq 0 \\ -\lambda(-g)^{\alpha_L} & g < 0 \end{cases} \tag{11}$$

with $0 < \alpha_G, \alpha_L \leq 1, \lambda \geq 1$.[14] As for CPT, the parameters $\alpha_G$ and $\alpha_L$ determine the investor's risk aversion over gains and risk seeking over losses while loss aversion is measured by $\lambda$. The scaling parameter satisfies $0 \leq \beta \leq \min(\alpha_L, \alpha_G)$. The upper restriction on $\beta$ ensures the desired property discussed earlier that $|U(G, R)|$ is weakly decreasing in $R$ for a fixed $G$. A nonnegative $\beta$ is a participation constraint.

Tversky and Kahneman (1992) estimated the utility parameters as $\alpha_G = \alpha_L = 0.88$ and $\lambda = 2.25$. Since they were not concerned about intertemporal aspects, they did not estimate a discount rate nor did they consider scaling. However, for such a low level of risk aversion, the transversality condition is violated and the investor waits forever to realize any gain unless $\delta$ is nearly equal to the expected rate of return.[15] For $\alpha_G = \alpha_L = 0.88$ and $\delta = 8\%$, the investor never voluntarily realizes losses unless there is little loss aversion with $\lambda$ close to one.

However, voluntary loss taking can be part of the optimal policy for other utility parameters. Wu and Gonzalez (1996), for example, estimate $\alpha = 0.5$. Using utility parameters, $\alpha_G = \alpha_L = 0.5$, $\lambda = 2$, and $\delta = 5\%$ the optimal strategy does include voluntary losses for any $\beta$ less than about 0.327.

Figure 2 shows the optimal sales strategies, $\Theta$ and $\theta$, plotted against $\lambda$ for different values of $\alpha_G$ and $\alpha_L$. Both $\Theta$ and $\theta$ decrease with $\lambda$ though $\theta$ falls at a much faster rate, and for a large enough loss aversion, the investor refrains from realizing any losses. Provided losses are realized, they are always larger than gains in magnitude. This might seem counterintuitive, but the smaller gains are realized more frequently, and since marginal utility is decreasing with the magnitude of the gain or loss, several small gains more than offset the disutility of a single loss of the same total size.

One common observation about the realization of gains and losses is the disposition effect, which has often been claimed to be a consequence of an S-shaped utility function.[16] The

---

[14] Setting $\beta = \alpha_G$ and $\lambda = \lambda_0 R^{\alpha_L - \alpha_G}$ reduces (11) to the standard case introduced in Tversky and Kahneman (1992), with $\lambda_0$ being their loss aversion parameter. Loss aversion then would vary with $R$, but the reference level is constant in the original static interpretation. The Barberis and Xiong (2012) model is the special case $\beta = \alpha_G = \alpha_L = 1$.

[15] The restriction on $\delta$ comes from the transversality condition, $\alpha_G \leq \gamma_1$. While the required discount rate is large relative to those usually assumed, many behavioral finance models do assume that investors are quite impatient. It seems reasonable that utility derived from trading gains might well display more impatience than utility for lifetime consumption. In addition, this high discount rate could incorporate the hazard rate describing the investor's ceasing this type of trading. Death is sometimes inserted into infinite-horizon models in this fashion, though here the termination of trading might be a simple lack of further interest.

[16] Shefrin and Statman (1985) were the first to argue that an S-shaped utility function leads to the disposition effect. Similar arguments were made in Weber and Camerer (1998), Odean (1998), Grinblatt and Han (2005), and other theoretical and empirical papers.

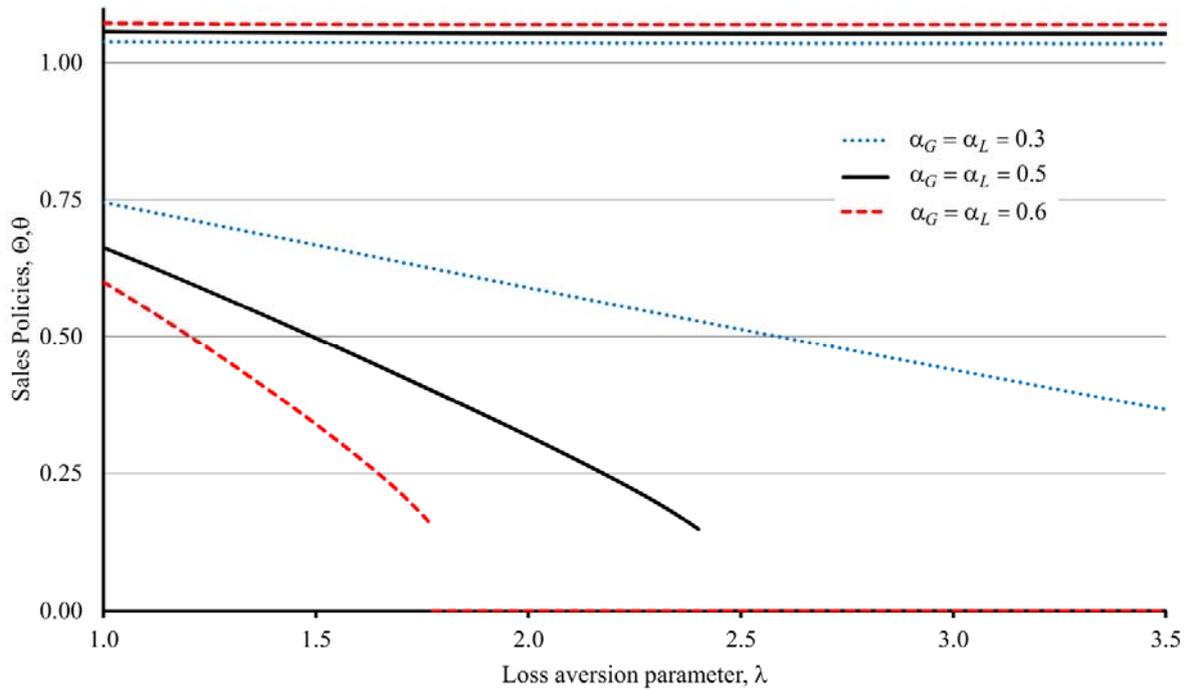

**Figure 2: Optimal Sales Policies.** This figure illustrates the optimal policies for selling at gains or losses as a multiple or fraction of the reference level. The stock price parameters are $\mu = 9\%$, $\sigma = 30\%$. Transaction costs are $k_s = k_p = 1\%$. Utility parameters are $\delta = 5\%$, $\beta = 0.25$, and $\alpha_G = \alpha_L = 0.3, 0.5, 0.6$, as indicated. The investor fully recognizes transaction costs in assessing his utility, i.e., $\kappa = K \equiv (1-k_s)/(1+k_p)$.

---

effect, which has often been claimed to be a consequence of an S-shaped utility function.[17] The reasoning is that the investor realizes his gains as he is risk averse and therefore unwilling to gamble about future uncertain gains; however, being risk-seeking over losses he will gamble and postpone realizing them. This analysis implicitly assumes something like realization utility because any effects of the unrealized gains or losses are ignored. But the argument is static considering only a single sale and ignoring any effects of reinvestment, nor does it address the question of why any losses are ever realized rather than their being continually postponed. Of course, even ignoring reinvestment, the realization of gains might be postponed if the expected change in the stock price is sufficiently high so that a larger expected gain in the future offsets its extra risk. Conversely, if the mean price change is negative, losses might be realized early to avoid larger expected losses in the future while gains would be realized both to avoid risk and to avoid smaller expected gains.[18]

---

[17] Shefrin and Statman (1985) were the first to argue that an S-shaped utility function leads to the disposition effect. Similar arguments were made in Weber and Camerer (1998), Odean (1998), Grinblatt and Han (2005), and other theoretical and empirical papers.

[18] Henderson (2012) formalizes this argument by examining a diffusion model like ours that allows only a single

With concave realization utility, the disutility of a loss can never be offset by the benefit of recovering that loss in subsequent gains of the same total size, but as seen in Figure 2 losses will be realized as well as gains with an S-shaped utility function. Losses are substantially less common than gains since θ is much farther from 1 than is Θ.[19] However, in sharp contrast with the static argument that an S-shaped utility leads to the disposition effect, the S-shape actually serves to reduce the disposition effect in a dynamic context. As $\alpha_G$ and $\alpha_L$ decrease and the S-shape becomes more pronounced, the optimal gain point, Θ, is affected only a little while the loss point, θ, increases dramatically, reducing the disposition effect. The reason is that realizing a loss resets the reference level for future possible gains, and this can more than offset the direct disutility of the loss. That is, the realization of a loss is, in some sense, the purchase of a valuable option. When $\alpha_G$ is small, this option effect can be substantial since the marginal utility of small gains is very large making the disutility of losses "affordable."

With intertemporal realization utility, an S-shaped utility function does not create the disposition effect; it actually reduces it, explaining why any voluntary losses are realized rather than none.[20]

As the loss aversion parameter λ increases, losses become more painful, and θ drops discontinuously to zero as shown in Figure 2. The discontinuous change occurs because this maximization problem is not a standard convex optimization. As illustrated in Figure 3, the reduced value function, $v(1)$, is not a concave function of θ. Both an interior local maximum and a corner local maximum at zero are possible and either can be the global maximum. The high marginal disutility of repeated small losses together with transaction costs makes loss taking suboptimal for high values of θ near 1. On the other hand for low θ, the continuation value, which is proportional to $(K\theta)^\beta$, is very small and cannot offset the disutility of a loss. This makes avoiding losses altogether (θ = 0) better than taking a large loss. Only for intermediate values of θ is the continuation value possibly sufficient to offset the disutility of a loss. So $v(1)$ attains its local maximum value at either θ = 0 or an intermediate value.

Figure 3 illustrates both optimum types for $\alpha_G = \alpha_L = 0.5$ and β = 0.3. The initial reduced value function after any sale and repurchase, $v(1)$, is plotted against θ for three values of λ. The upper sales point is fixed at its distinct optimal value in each case. For λ = 2.5, it is optimal to sell for a loss at θ = 0.183. For λ = 2.56, there is an optimum at θ = 0.147, but this is only a local maximum as never selling at a loss provides higher utility as shown. For the critical value of λ ≈ 2.531, both selling for a loss at θ = 0.166 and never selling for a loss provide the same expected utility. Therefore, the lower sales point, θ, does not decrease smoothly to zero as λ increases; it

---

liquidating sale with no reinvestment. She finds that losses are voluntarily realized only if μ < 0. In contrast to her model, our paper shows that reinvestment is important, and as a direct consequence, there is voluntary realization of losses even with a positive μ of empirically relevant magnitude.

[19] Since returns are lognormal, the proper "distance" comparison is $\ell n \Theta \ll -\ell n \theta$; also $\mu - \frac{1}{2}\sigma^2 > 0$, so even if the log distances were equal, gains would be realized more often.

[20] Some research suggests that prospect theory may not lead to the disposition effect, e.g., Barberis and Xiong (2009), Kaustia (2010), and Hens and Vlcek (2011). In contrast with our model, none of these papers consider reinvestment.

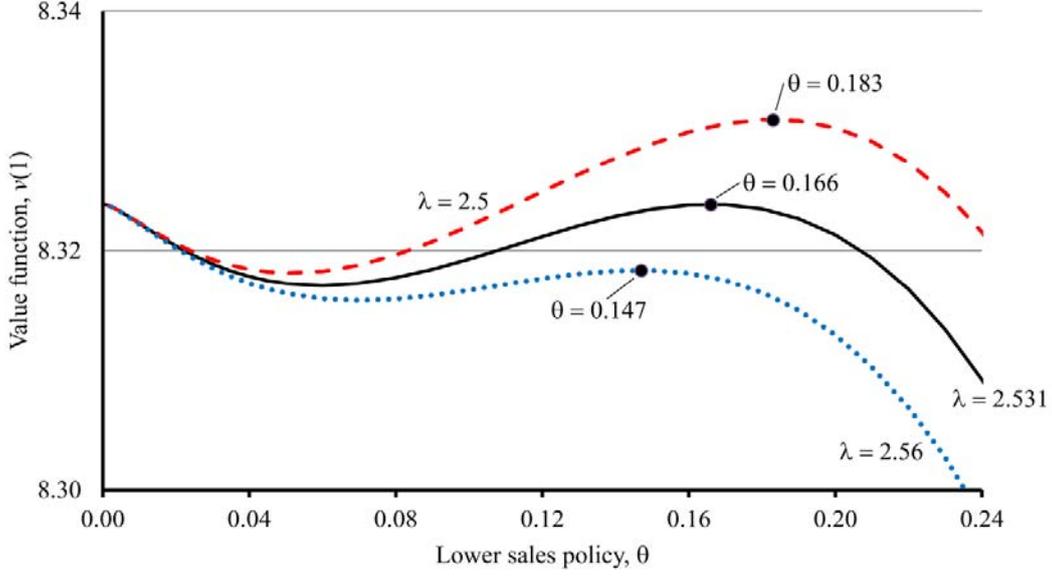

**Figure 3: The Value Function for Scaled-TK Utility.** The reduced value function for scaled-TK utility measured immediately after a sale and reference level reset, $v(1)$, is plotted against different loss sales points, $\theta$. The gain sales point, $\Theta$, is fixed at its optimal value. The solid line shows the value function for $\lambda = 2.531$, the dashed line shows the value function for $\lambda = 2.5$, and the dotted line shows the value function for $\lambda = 2.56$. The other parameters are $\mu = 9\%$, $\sigma = 30\%$, $k_s = k_p = 1\%$, $\alpha_G = \alpha_L = 0.5$, $\beta = 0.3$, $\delta = 5\%$ $\kappa = 1-k_s$. For $\lambda = 2.5$, the two-point policy ($\theta = 0.183$, $\Theta = 1.037$) is optimal. For $\lambda = 2.56$, the one-point policy ($\Theta = 1.036$) is optimal. For the critical value $\lambda = \lambda_* = 2.531$, the two point policy ($\theta = 0.166$, $\Theta = 1.036$) and the one-point policy ($\Theta = 1.036$) have the same expected utility.

---

drops discontinuously from 0.166 to 0 as $\lambda$ passes the critical value of 2.531. A similar change in regime for $\theta$ is true for the other parameters. The two regimes are characterized in Proposition 1; a proof is supplied in the Appendix.

**Proposition 1:** Scaled-TK utility has both an upper and a (non-zero) lower optimal sales point if and only if $\lambda$ is less than the critical value

$$\lambda_* = \frac{(\kappa\Theta_* - 1)^{\alpha_G - 1}\theta_*^\beta}{(1 - \kappa\theta_*)^{\alpha_L - 1}\Theta_*^\beta} \times \frac{(\alpha_G - \gamma_1)\kappa\Theta_* + \gamma_1}{(\alpha_L - \gamma_1)\kappa\theta_* + \gamma_1}$$

where (12)

$$0 = (\alpha_G - \gamma_1)\kappa\Theta_*^{\gamma_1+1-\beta} + \gamma_1\Theta_*^{\gamma_1-\beta} - (\alpha_G - \beta)K^\beta\kappa\Theta_* - \beta K^\beta$$

$$0 = (\alpha_L - \gamma_1)\kappa\theta_*^{\gamma_1+1-\beta} + \gamma_1\theta_*^{\gamma_1-\beta} - (\alpha_L - \beta)K^\beta\kappa\theta_* - \beta K^\beta$$

determine $\Theta_*$ and $\theta_*$. If $\lambda$ is greater than this critical value, only gains are realized. The solution is still characterized by (8), (9), and (10) with $C_2$ set to 0. As $\beta$ approaches its transversality upper limit, $\gamma_1$, $\lambda_* \to 0$, and voluntary losses are never realized. ∎

The Barberis Xiong (2012) model is a special case of scaled-TK utility with $\alpha_L = \alpha_G = \beta = 1$. For this model, or indeed any realization utility model with piece-wise linear utility for gains and losses and $0 \leq \beta \leq \gamma_1$, the critical value $\lambda_*$ is less than 1. Therefore, losses are never realized voluntarily.

## 2. The Value Function and Optimal Sales Policies

The value function or its reduced-form equivalent, $v$, measures the present value of the investor's utility bursts and gives a point estimate of the benefit of his strategy. It serves the role of the derived utility function in a standard Merton-type portfolio problem.

Figure 4 presents the reduced value function, $v$, measured at the time of any reinvestment, i.e., $v(1)$, plotted against the asset's expected rate of return, $\mu$, and standard deviation, $\sigma$. The default utility parameters are $\lambda = 2$, $\delta = 5\%$, $\beta = 0.3$, $\alpha_G = \alpha_L = 0.5$. For the $\mu$ and $\sigma$ graphs, the other parameter is set to $\sigma = 30\%$ or $\mu = 9\%$, respectively. For comparison purposes, each value

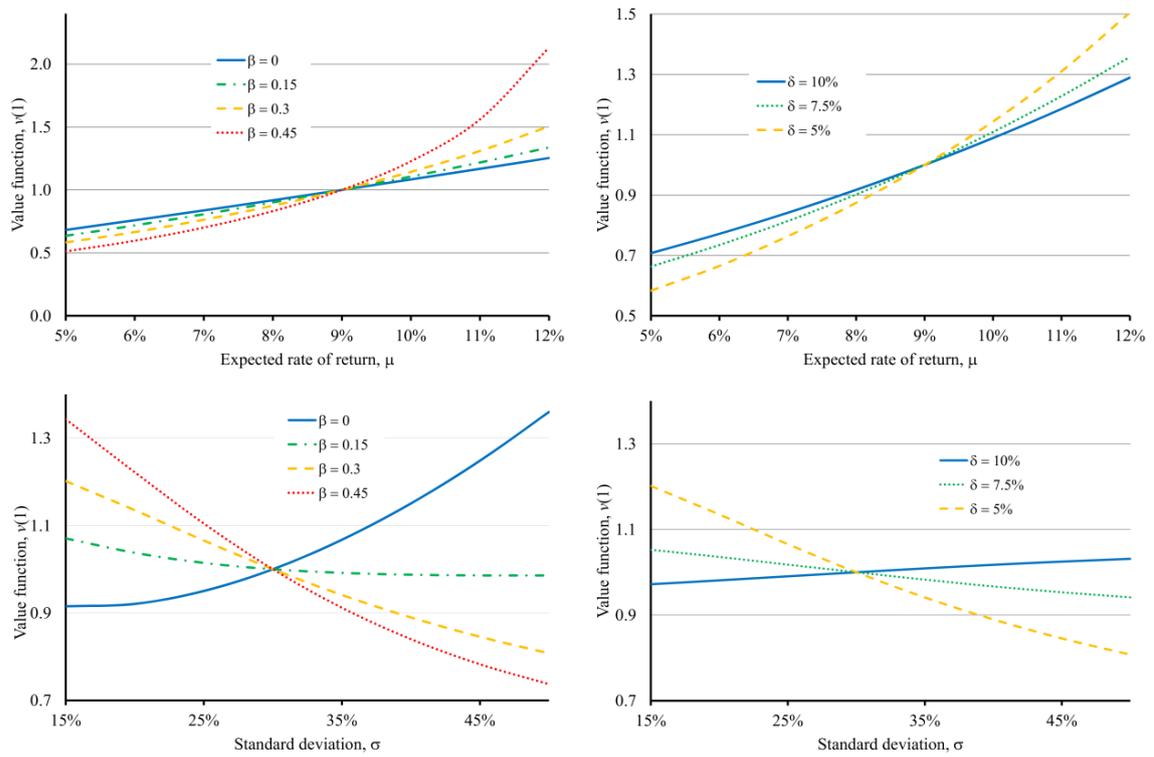

**Figure 4: The Value Function.** The initial optimized value function for scaled-TK utility plotted against $\mu$ and $\sigma$. The default parameters are $\mu = 9\%$, $\sigma = 30\%$, $k_s = k_p = 1\%$, $\beta = 0.3$, $\lambda = 2$, $\delta = 5\%$, $\alpha_G = \alpha_L = 0.5$. The investor fully recognizes transaction costs in assessing his reinvestment gains, i.e., $\kappa = K \equiv (1-k_s)/(1+k_p)$.

function is normalized to 1 at the values $\mu = 9\%$ and $\sigma = 30\%$.[21]

Naturally the value function is increasing in $\mu$. On average for higher $\mu$, the next trade is more likely to be a gain and to occur sooner. The relation is steeper for a larger $\beta$ because the continuation value from the reinvestment is larger due to scaling. The relation is also steeper for smaller $\delta$ since the benefits from future gains are discounted less heavily.

Surprisingly, the value function is not always strictly decreasing in volatility as it is for a standard expected utility maximization; it can be increasing or U-shaped. Of course, CPT investors are risk-seeking with respect to losses, but that is not the reason for this effect. For example, the value function is increasing in volatility in the Barberis and Xiong (2012) model where burst utility is piece-wise linear and weakly concave. In our model there are conflicting effects.

Changing the three parameters, $\mu$, $\delta$, and $\sigma^2$ proportionally is identical to a change in the unit of time and leaves our model unaffected. So an increase in $\sigma^2$ can be interpreted as a proportional decrease in both $\mu$ and $\delta$. Decreasing $\mu$ lowers the value function as just explained, but decreasing $\delta$ raises the value function since the future net positive bursts are discounted less heavily. As explained above, the smaller is $\beta$, the less important is the $\mu$ effect. So for small $\beta$, the value function is less steeply decreasing or even increasing in volatility. Also the larger is $\delta$ the more important is its effect. So for large $\delta$, the value function is again less steeply decreasing or even increasing in volatility.

Figure 5 presents graphs of the optimal selling points for gains and losses for scaled-TK investors. The parameters left unchanged in each graph are set to the default values $\mu = 9\%$, $\sigma = 30\%$, $k_s = k_p = 1\%$, $\alpha_G = \alpha_L = 0.5$, $\beta = 0.3$, $\lambda = 2$, $\delta = 5\%$. The dotted lines show the optimal policies for an investor who ignores the reinvestment cost in assessing his gains, i.e., $\kappa = 1-k_s$. The solid lines show the optimal policies for an investor who does recognize this cost, i.e., $\kappa = (1-k_s)/(1+k_p)$. Several features are immediately evident.

For both types of investors, realized losses are typically much larger than realized gains so the basic strategy is to realize a few large losses and many small gains as we have already suggested intuitively. However, the no-sales region is wider for an investor who recognizes the reinvestment cost as reducing his gain. An investor who internalizes the costs more when assessing his well-being is obviously more reluctant to trade.

The upper sales point, $\Theta$, is much less affected by parameter changes than is the lower sales point, $\theta$, in most cases. In fact, $\Theta$ is largely unaffected by any of the variables except transaction costs and the scaling parameter, $\beta$. And for $\beta$, any effect occurs mostly near the transversality limit. As $\beta$ approaches its limit of $\gamma_1$, $\theta$ drops discontinuously to zero, and $\Theta$ approaches $\beta/[\kappa(\beta-\alpha_G)]$.[22]

---

[21] Standard utility functions are defined only up to a positive affine transformation. Realization utility has its level set so that a gain of zero gives a utility of zero, but scaling is still arbitrary.

[22] For $\beta > \gamma_1$, there is no well-defined optimal upper sales point. Any $\Theta > K^{-\beta/(\beta-\gamma_1)}$ provides infinite expected utility. See the Appendix for details on this and other transversality-type violations. When $\alpha_G$ is very close to 1, the upper sales point is sensitive to $\beta$ and can be decreasing for low $\beta$.

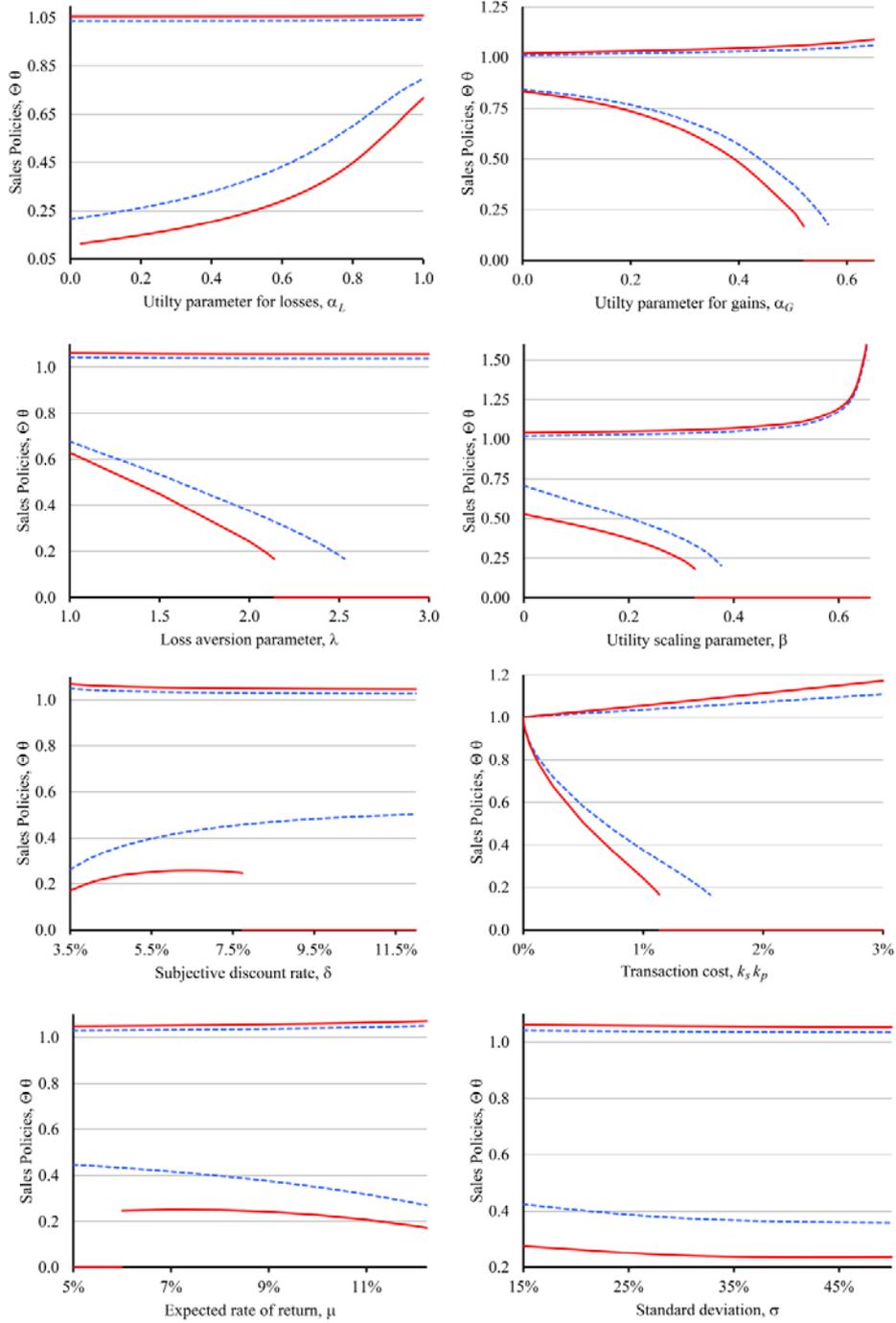

**Figure 5: The Optimal Sales Policies for Scaled-TK Utility.** The optimal sales points, $\Theta$ and $\theta$, for scaled-TK utility plotted against various parameters. The default parameters in each graph are $\mu = 9\%$, $\sigma = 30\%$, $k_s = k_p = 1\%$, $\alpha_G = \alpha_L = 0.5$, $\beta = 0.3$, $\lambda = 2$, $\delta = 5\%$. The dotted lines show the optimal policies for an investor who ignores the reinvestment cost in assessing his gains, $\kappa = 1-k_s$. The solid lines show the optimal policies for an investor who recognizes the reinvestment cost in assessing his gains, $\kappa = K \equiv (1-k_s)/(1+k_p)$.

On the other hand, the optimal loss-taking strategy does vary substantially as the utility parameters change. Increasing $\lambda$ makes losses hurt more so $\theta$ is lowered in response to avoid some of them. Smaller $\alpha_L$ leads to higher marginal disutility for small losses. This induces the investor to wait longer to realize a loss as his risk-seeking behavior increases ($\alpha_L$ decreases from 1 to 0). Conversely for low $\alpha_G$, the marginal utility of small gains is quite high making small losses "affordable" and desirable to set up future gains. So $\theta$ decreases with $\alpha_G$. Similarly when $\beta$ is small, the pain of realizing a loss is offset more by the lowering of the reference level for subsequent gains; this intensifies loss taking, increasing $\theta$. With a higher $\beta$, the investor is more reluctant to experience a loss just to increase future gains.

The effect of the subjective discount rate, $\delta$, on the optimal sales policy is unusual. A more impatient investor wants to realize gains sooner and defer losses longer. We see that $\Theta$ is decreasing in $\delta$ as expected; however, $\theta$ is not. All losses are taken voluntarily, and the desire to take gains early induces a derived willingness to realize losses in order to set up these future gains. This causes $\theta$ to also be increasing in $\delta$ at low discount rates; however, at higher discount rates the initial intuition dominates because future gains are discounted more so the benefit of resetting is less.

Increasing the transaction costs, $k_s$ and $k_p$, naturally widens the no-sales region because the costs take part of each gain and increase every loss. As the costs go to zero, both $\Theta$ and $\theta$ approach 1, and the trading frequency increases without limit. Because marginal utility becomes infinite as the gain size goes to zero, the investor takes every opportunity to realize even the smallest of gains. Of course, there is unbounded marginal disutility for near-zero losses, but under the optimal strategy $\theta$ approaches 1 slower than does $\Theta$, so there is a net increase in the value function with frequent trades.

There is a similar result for any utility function that is strictly concave for gains even if their marginal utility is not infinite at zero. There is always an incentive to realize any gain as a series of smaller increments because the marginal utility is highest near 0, and in the absence of transactions coasts, there is no selling penalty to offset this. However, when the marginal utility for gains is not infinite at zero, the loss sales point, $\theta$, need not be near 1. In the presence of loss aversion ($\lambda > 1$), the marginal utility of small losses exceeds the marginal utility of small gains which precludes an immediate loss realization even in the absence of transaction costs.

Changes in $\mu$ have very little effect on the size of optimal realized gains, $\Theta$. The value function is strongly increasing in $\mu$, but this is due to the reduction in the average time before a sale occurs rather than any significant change in policy. The lower sales point, $\theta$, is affected more. For large or small $\mu$, the option to reset the reference point by selling at a loss is less valuable than for intermediate values of $\mu$. It is less valuable for large $\mu$ because the asset price grows quickly enough for gains to be realized without a painful reset. Conversely, with a very low $\mu$, there is less value in resetting the reference level since future gains will be realized infrequently. So $\theta$ is highest for intermediate values of $\mu$.[23]

---

[23] Note that this is very different from the result in Henderson (2012). In her model with neither reinvestment nor discounting, losses are only realized to avoid even larger losses on average when $\mu$ is negative.

The standard deviation, σ, has only a tiny impact on the optimal gain point. Θ decreases as σ rises but by amount imperceptible in the graphs. Increasing σ also lowers the loss sales point, θ. The reason is, that for typical parameter values, higher volatility would increase the probability of loss realization, and the investors responds by lowering θ to postpone voluntary loss taking.

## 3. Model Calibration

In this section we calibrate our model using representative investors to determine if it can explain observed trading patterns. The model described in the previous sections makes specific predictions about the magnitudes, frequencies, and relative proportion of both realized and unrealized gains and losses. In this context we explore an alternate utility specification which improves the model predictions. We also compare our model results to predictions assuming random trading. The best calibration is achieved when we consider heterogeneous trading strategies.

If we consider a single set of utility and stock price parameters, our model predicts that the magnitudes of all realized gains and losses are $\Theta-1$ and $\theta-1$, respectively. The frequency of trading is determined by the time required for the investment value to rise to $\Theta R$ or fall to $\theta R$ from the original reference level $R$. Paper gains expressed as a percentage of the reference level are distributed over the range $\theta-1$ to $\Theta-1$. The properties of these distributions are given in Proposition 2. Its proof and that of all later propositions are provided in the Appendix.

**Proposition 2: Properties of Realized and Unrealized Gains and Losses.** If the asset value has a lognormal evolution, $dX/X = \mu dt + \sigma d\omega$, and the investor trades repeatedly according to a constant two-point policy, θ-Θ, then the probabilities that a given episode eventually ends with a gain or a loss are[24]

$$Q_G = \frac{\theta^{-\eta}-1}{(\Theta/\theta)^\eta - 1} \qquad Q_L = 1 - Q_G = \frac{1-\Theta^{-\eta}}{1-(\theta/\Theta)^\eta} \qquad (13)$$

$$\text{where} \quad \eta \equiv 1 - 2\mu/\sigma^2 \,.$$

The fractions of the time the asset has an unrealized gain or loss are

$$\varphi_G = \frac{(1-\theta^\eta)[\ell n\, \Theta - \tfrac{1}{\eta}(\Theta^\eta - 1)]}{(1-\theta^\eta)\ell n\, \Theta + (\Theta^\eta - 1)\ell n\, \theta} \qquad \varphi_L = 1 - \varphi_G \,. \qquad (14)$$

The expected duration of each investment episode is

$$\mathbb{E}[\tau] = \frac{(\Theta^\eta - 1)\ell n\, \theta - (\theta^\eta - 1)\ell n\, \Theta}{(\Theta^\eta - \theta^\eta)(\mu - \tfrac{1}{2}\sigma^2)} \,, \qquad (15)$$

---

[24] When $2\mu = \sigma^2$ (so $\eta = 0$), L'Hôspital's rule gives $Q_G = 1 - Q_L = \varphi_L = 1 - \varphi_G = -\ell n\, \theta/\ell n(\Theta/\theta)$ and $\mathbb{E}[\tau] = -\ell n(\theta)\ell n(\Theta)/\sigma^2$.

and in a sequence of consecutive investments, the expected number of investment episodes per unit time is $1/\mathbb{E}[\tau]$. ∎

The properties derived in this proposition and Proposition 4 below do not depend on the specific realization utility function assumed nor even on the maximization of utility at all. They obtain whenever a specific two sales point strategy is employed for assets with lognormal diffusions.

For comparison we want the same type of predictions for investors who may trade for liquidity purposes, based on information, or for other reasons. Describing the actions of all such investors is outside the scope of this paper so we simply assume that these investors trade stocks randomly in separate episodes with each episode terminated independently of the stock price evolution. The predictions of this model are given in Proposition 3.

**Proposition 3: Characteristics of Investment Episodes for Random Trades**. Assume that each asset's price evolves according to a lognormal diffusion and that each trading episode terminates with a sale that occurs according to a Poisson process which is independent of the evolution of the stock price and has intensity $\rho$. The trading episodes have the properties given below.

The duration of each episode has an exponential distribution with mean duration $\mathbb{E}[\tau] = 1/\rho$. The average realized gain and loss are[25]

$$\overline{\Theta} = -\frac{\rho(1-\psi^-)}{(\rho-\mu)\psi^-} \qquad \overline{\theta} = \frac{\rho(\psi^+ - 1)}{(\rho-\mu)\psi^+} \tag{16}$$

$$\text{where} \quad \psi^{\pm} \equiv \frac{-(\mu - \tfrac{1}{2}\sigma^2) \pm \sqrt{(\mu - \tfrac{1}{2}\sigma^2)^2 + 2\rho\sigma^2}}{\sigma^2}.$$

The probability that a given episode will end with a sale at a gain and the probability that an unrealized investment is a paper gain are both

$$Q_G = \varphi_G = \frac{\psi^-}{\psi^- - \psi^+}. \tag{17}$$

Of course, the probability that a given episode will end in a loss and the probability of an unrealized loss are $Q_L = \varphi_L = 1 - Q_G = 1 - \varphi_G$. ∎

Note that the trading points, $\overline{\Theta}$ and $\overline{\theta}$, given in (16) are averages. While threshold-realization-utility investors always trade at fixed ratios, the sales of random traders can occur at any price.

---

[25] For $\rho = \mu$, L'Hôspital's rule gives $\overline{\theta} = \mu/(\mu + \tfrac{1}{2}\sigma^2)$. The expected value of the upper sales, $\overline{\Theta}$, is finite only if $\rho > \mu$; $\overline{\theta}$ is always finite since its realizations are bounded between 0 and 1.

One set of trading statistics that has garnered considerable attention was proposed by Odean (1998) to measure the disposition effect. These include the proportion of gains realized, *PGR*; the proportion of losses realized, *PLR*; and the Odean measure, $\mathcal{O}$. *PGR* is defined as a ratio. The numerator is the number of realized gains summed over all days and all accounts. The denominator is the total over all days of the number of stock positions showing a gain (realized or not) in all accounts which realized either a gain or a loss on that day. *PLR* is similarly defined for losses. The Odean measure is the ratio *PGR*/*PLR*.

In a given sample, these statistics will be affected by many factors. These include the varying sales thresholds for the distinct assets held by different investors, the number of stocks held in each account, the correlation between the stocks' returns, and the sampling interval.[26] Proposition 4 derives *PGR*, *PLR*, and the Odean measure for the special case of independent and identically distributed assets. The number of stocks held per account may vary across investors, but each individual stock is traded according to the same two-point or random strategy.

**Proposition 4: Odean's Statistics with a Representative Investor.** Assume that asset returns are independent and identically distributed with a lognormal evolution, $dX/X = \mu dt + \sigma d\omega$, and that all stocks are traded using the same strategy. Then as the number of sales increases, the probability limits of *PGR*, *PLR*, and the Odean measure are

$$\text{plim}\, PGR = \frac{Q_G}{Q_G + (\bar{n} + \sigma_n^2/\bar{n} - 1)\varphi_G} \quad \text{plim}\, PLR = \frac{Q_L}{Q_L + (\bar{n} + \sigma_n^2/\bar{n} - 1)\varphi_L}$$

$$\text{plim}\, \mathcal{O} = \frac{\text{plim}\, PGR}{\text{plim}\, PLR} = \frac{Q_G}{Q_L} \frac{Q_L + (\bar{n} + \sigma_n^2/\bar{n} - 1)\varphi_L}{Q_G + (\bar{n} + \sigma_n^2/\bar{n} - 1)\varphi_G}$$

(18)

where $\bar{n}$ and $\sigma_n^2$ are the average number and variance of the number of stocks held across accounts,[27] and $Q_G$, $Q_L$, $\varphi_G$, and $\varphi_L$ are the probabilities defined in (13) and (14) if the investors are realization-utility investors or in (17) if the investors are random Poisson traders. ∎

The statistics of Proposition 4 are the same as those that would be produced by a single representative investor holding $\bar{n} + \sigma_n^2/\bar{n}$ rather than $\bar{n}$ stocks. The representative investor's holding is biased high relative to the average because those investors holding more stocks are represented more often in the data.

With the statistics derived in Propositions 2 through 4, we can assess how our realization utility model and the random trading model fit the trading patterns of individual traders. For comparison we incorporate data from Odean (1998) and Dhar and Zhu (2006) with the statistics generated by this model into Table 1. The first row of the table presents Odean's data extracted from his Tables 1 and 3 including the header text. He reports that 53.8% of sales were realized

---

[26] The effects of these factors are examined in Ingersoll and Jin (2012).

[27] If each account holds a fixed number of assets over time, then $\bar{n}$ and $\sigma_n^2$ are the average and variance of account sizes. Under mild regularity conditions, the proposition remains valid for accounts whose sizes vary over time with $\bar{n}$ and $\sigma_n^2$ being the average and variance of the number of shares held per account across both accounts and time.

**Table 1: Summary Statistics for Reference-Dependent Realization Utility Model with Scaled Tversky-Kahneman Utility**

The table reports: $\Theta-1$, $\theta-1$: percentages above and below the reference level for realized gains and losses, $Q_G$: fraction of episodes that end in realized gains, $\varphi_G$: fraction of stocks with unrealized paper gains, $\mathbb{E}[\tau]$: average holding period in trading days (250 per year), *PGR*, *PLR*: proportions of gains and losses realized, and $\mathcal{O} \equiv PGR/PLR$: Odean's measure. Asset parameters are $\mu$ = 9% and $\sigma$ = 30%. The accounts' sizes are fixed with $\bar{n} + \sigma_n^2/\bar{n} = 8.0$. Utility parameters are $\lambda$ = 2 and $\delta$ = 5% (except $\delta$ = 8% for $\alpha_G$ = 0.88 and $\delta$ = 10% for $\alpha_G$ = 1 to avoid a transversality violation). Transaction costs are $k_s = k_p$ = 1%, and the investor accounts for both costs in his subjective view of his realized gains, $\kappa = K$. Odean's data is taken from Tables 1 and 3 of his 1998 paper. Dhar and Zhu's data is from the notes to their Table 3. The "Fit to Odean's $\Theta$, $\theta$" row uses Odean's estimates of $\Theta$ and $\theta$ to compute the other values using Propositions 2 and 4. Each "Poisson Model" row chooses $\rho$ to match one of Odean's estimates of $\Theta$, $\theta$, $Q_G$, or $\mathbb{E}[\tau]$ and computes the other values using Propositions 3 and 4; the observed $\varphi_G$ cannot be matched as the Poisson model cannot give values less than 50% when $\mu > \sigma^2/2$.

| | | $\Theta-1$ | $\theta-1$ | $Q_G$ | $\varphi_G$ | $\mathbb{E}[\tau]$ | *PGR* | *PLR* | $\mathcal{O}$ |
|---|---|---|---|---|---|---|---|---|---|
| | Odean data | 27.7% | −22.8% | 53.8% | 41.9% | 312 | 14.8% | 9.8% | 1.51 |
| | Dhar & Zhu data | ---- | ---- | 65.8% | 46.5% | 122 | 13.2% | 6.4% | 2.06 |
| | Fit to Odean's $\Theta$, $\theta$ | **27.7%** | **−22.8%** | 57.7% | 50.7% | 174 | 14.0% | 10.9% | 1.28 |

| Random Trading (Poisson) Model | | | | | | | | | |
|---|---|---|---|---|---|---|---|---|---|
| | | $\Theta-1$ | $\theta-1$ | $Q_G$ | $\varphi_G$ | $\mathbb{E}[\tau]$ | *PGR* | *PLR* | $\mathcal{O}$ |
| | $\rho = 0.36$ | 72.2% | **−22.8%** | 58.7% | 58.7% | 688 | 12.5% | 12.5% | 1 |
| | $\rho = 0.80$ | 36.4% | −17.4% | 55.9% | 55.9% | **312** | 12.5% | 12.5% | 1 |
| | $\rho = 1.16$ | **27.7%** | −15.2% | 54.9% | 54.9% | 215 | 12.5% | 12.5% | 1 |
| | $\rho = 1.94$ | 19.7% | −12.4% | **53.8%** | 53.8% | 129 | 12.5% | 12.5% | 1 |

| Realization Model with Scaled-TK Utility | | | | | | | | | |
|---|---|---|---|---|---|---|---|---|---|
| | | $\Theta-1$ | $\theta-1$ | $Q_G$ | $\varphi_G$ | $\mathbb{E}[\tau]$ | *PGR* | *PLR* | $\mathcal{O}$ |
| $\alpha_G = 1$ | $\beta = 0$ or $1$ | 95.3% | never | 100% | 27.1% | 3717 | 34.5% | 0 | $\infty$ |
| $\alpha_L = 1$ | $\beta = 0.53$ | 45.6% | never | 100% | 16.6% | 2087 | 46.2% | 0 | $\infty$ |
| $\alpha_G = 0.88$ | $\beta = 0$ | 17.6% | never | 100% | 7.7% | 901 | 65.0% | 0 | $\infty$ |
| $\alpha_L = 0.88$ | $\beta = 0.88$ | 96.2% | never | 100% | 27.3% | 3743 | 34.4% | 0 | $\infty$ |
| $\alpha_G = 0.5$ | $\beta = 0$ | 3.9% | −13.5% | 80.6% | 21.5% | 15 | 34.9% | 3.4% | 10.22 |
| $\alpha_L = 0.88$ | $\beta = 0.3$ | 5.8% | −45.3% | 93.8% | 9.5% | 85 | 58.6% | 1.0% | 60.64 |
| $\alpha_G = 0.5$ | $\beta = 0$ | 3.8% | −6.3% | 64.9% | 36.7% | 7 | 20.2% | 7.3% | 2.74 |
| $\alpha_L = 1.0$ | $\beta = 0.3$ | 5.9% | −28.2% | 87.6% | 15.6% | 50 | 44.5% | 2.1% | 21.64 |
| $\alpha_G = 0.5$ | $\beta = 0$ | 4.0% | −47.3% | 95.9% | 6.5% | 63 | 67.8% | 0.6% | 107.90 |
| $\alpha_L = 0.5$ | $\beta = 0.3$ | 5.7% | −75.8% | 98.3% | 4.9% | 169 | 74.3% | 0.3% | 293.06 |

gains with an average size of 27.7%; the remaining trades were losses averaging –22.8%. The average holding period was 15 months which we have expressed as 312 trading days.[28] Paper gains and losses composed 41.9% and 58.1% of the unrealized positions. *PGR* and *PLR* were 14.8% and 9.8%. Dhar and Zhu's (2006) data are taken from their Table 1 and the note to their Table 3. Gains were realized on 65.8% of trades, but paper gains composed only 46.5% of unrealized positions. *PGR* and *PLR* were 13.2% and 6.4%.[29] The average holding period was 122 days. They do not report the average sizes of realized gains and losses. The differences in this data can probably be attributed to the periods studied. During the Dhar-Zhu period, 1991–6, the market rose 113% with only minor corrections while during Odean's period 1987–93, the market rose only 89% and suffered two major downturns. So Dhar-Zhu traders would have reached their $\Theta$-points more frequently while Odean's traders would have had more opportunities to sell at losses.

To determine if any calibration is feasible, the data in the third row uses just Odean's average sales price ratios as estimates for $\Theta$ and $\theta$. The remaining values are determined from them and the stock evolution parameters using Propositions 2 and 4. This fit is not optimized; we have simply chosen an asset comparable to a typical share of stock with $\mu = 9\%$ and $\sigma = 30\%$. The fit for $Q_G$ and $\varphi_G$, and therefore the corresponding loss statistics, do seem reasonable allowing for sampling error and heterogeneity of assets and investors in the actual sample.[30] To compute *PGR*, *PLR*, and $\mathcal{O}$, we need account size information. Goetzmann and Kumar (2008), using the same data set as in Barber and Odean (2000), provide more details about portfolio sizes. They give the percentages of accounts of various sizes in their Table 1 from which we compute approximate values of $\bar{n} = 4.1$ and $\sigma_n = 4.0$ giving $\bar{n} + \sigma_n^2/\bar{n} \approx 8.0$. For a similar data set, Barber and Odean (2000) report that the average number of stocks per account is 4.3; Dhar and Zhu (2006) give average account sizes of 4.4 and 4.2 for investors whose occupations they identify as professional and nonprofessional.

The next four rows of the table illustrate the fit of a trading model based on random Poisson trades to Odean's data. As $\rho$ increases, both average sales points, $\overline{\Theta}$ and $\overline{\theta}$, approach 1. Under random trading, $Q_G$ and $\varphi_G$ must be equal, and both fall from 100% to 50% as $\rho$ increases from 0 to $\infty$.[31] While the individual statistics can be matched, they cannot be fit simultaneously.

---

[28] This calculation assumes 15 months is an exact figure of 312.5 days. The actual value could range from 302 to 323 days due to rounding.

[29] These are reported in their note to Table 3 using Odean's method of aggregation. In their Table 2, Dhar and Zhu (2006) report simple averages across investors for *PGR* and *PLR* of 38% and 17%, respectively. Computing *PGR* and *PLR* first at the investor level and then averaging across investors puts relatively more weight on investors who have fewer stocks in their accounts, and these investors typically have higher *PGR* and *PLR*. For instance, suppose $Q_G = \varphi_G = 0.5$. Then for an equal mix of investors who hold 2 and 6 stocks, *PGR* is 0.5 and 0.167. The average *PGR* is 0.33, but using equation (18) with $\bar{n} = 4$ and $\sigma_n = 2$, the aggregated *PGR* is 0.2.

[30] Our fitted value of 174 days for $\mathbb{E}[\tau]$ differs from both Odean's and Dhar-Zhu's though it is between them. All of the statistics in the last six columns except for $\mathbb{E}[\tau]$ depend only on the ratio $\mu/\sigma^2$ so increasing $\mu$ and $\sigma^2$ proportionally will reduce $\mathbb{E}[\tau]$ and leave the others unchanged. In our analysis below it is only the *relative* holding times for different accounts that matters not the level.

[31] If $2\mu < \sigma^2$, then both $Q_G$ and $\varphi_G$ rise from 0 to 50% as $\rho$ increases. In this case Odean's value of $Q_G = 53.8\%$ cannot be matched. Conversely, for the parameters used here, $\varphi_G$ cannot be matched to Odean's value of 41.9%.

In addition, both *PGR* and *PLR* must equal $(\bar{n} + \sigma_n^2/\bar{n})^{-1}$ with random trading so the Odean measure must always be 1.

The final ten rows of the table attempt to fit specific realization utility functions to Odean's data. Using $\alpha_G = \alpha_L = 0.88$ and $\lambda = 2.25$, as proposed by Tversky and Kahneman (1992), and $\delta = 8\%$, the upper sales point of our model varies from 17.6% to 96.2% above the reference level as the scaling parameter, $\beta$, ranges over its permitted values 0 to 0.88. This does include Odean's estimate of 27.7%, but the fit is far from satisfactory as no losses are voluntarily realized for these parameters.

For piecewise linear utility, $\alpha_G = \alpha_L = 1$ (and $\delta = 10\%$ to avoid transversality violation), the table confirms that voluntary loss taking is again never optimal as was shown in Proposition 1. Gains are realized after the stock price has risen by 95.3% for $\beta = 0$ or 1 or by a lesser amount for any $\beta$ between those values. The smallest gain-realization point is 45.6% when $\beta = 0.53$. It is apparent from the table that realization utility model cannot fit Odean's data with $\alpha$ values this high.

Tversky and Kahneman's estimates are from experimental settings with small gamble sizes. For the much larger size of investment that a typical investor makes in financial markets we expect more risk aversion.[32] Therefore, in Table 1 we also use $\alpha_G = 0.5$ which is the estimate of Wu and Gonzalez (1996).[33] This also permits lowering the rate of time preference to a more reasonable 5%. Since Wu and Gonzalez only estimate $\alpha_G$, we use $\alpha_L$ in the range 0.5 to 1.0.

From Table 1 it is apparent that the basic model can generate a wide variety of optimal sales points; however, for any parameter values that permit voluntary sales at the size of losses observed in Odean's data, the upper sales point, $\Theta$, is much too low. As a direct result, sales at gains vastly outnumber sales at loses ($Q_G \gg Q_L$) and *PGR* is too large while *PLR* is too small.

One difficulty with scaled-TK utility is that its derivative is very high near zero, indeed $u'(0) = \infty$ for any $\alpha_G < 1$. This makes the total utility of realizing gains in numerous tiny increments very large and pushes the optimal threshold, $\Theta$, quite close to 1. This is a particular problem in our model because sales, and therefore the sizes of any gains, are not exogenous but completely at the discretion of the investor.[34]

---

[32] Barber and Odean (2000) report that the average household in their sample holds 4.3 stocks worth $47,334, so the averaged dollar amount invested per stock a little more than $11,000. In Table 6 of Tversky and Kahneman (1992), the largest gamble about which subjects were questioned was $401, which represents a modest 3.6% gain or loss on the average stock position.

[33] Wu and Gonzalez (1996) only estimate $\alpha_G$. Their estimation depends on the form for the probability weighting function. When using that proposed in Tversky and Kahneman (1992), they estimate $\alpha_G = 0.5$; using the form proposed in Prelec (1998), they estimate of $\alpha_G = 0.48$.

[34] The Arrow-Pratt measure of absolute risk aversion for scaled-TK utility is $(1-\alpha_G)/G$ for uncertain prospects with gains only. So if agents have moderate risk aversion for moderately sized gains, they must be extremely risk averse about small gambles and close to risk-neutral for large ones. For modified-TK utility introduced below, the Arrow-Pratt measure is $(1-\alpha_G)/(R+G)$ which has less variation as the size of the gamble changes.

To avoid this problem, we consider a modified-TK utility function

$$U_{\text{mKT}}(G,R) = \begin{cases} R^{\beta}[(1+G/R)^{\alpha_G} - 1]/\alpha_G & G \geq 0 \\ -\lambda R^{\beta}[1 - (1+G/R)^{\alpha_L}]/\alpha_L & -R \leq G < 0. \end{cases} \quad (19)$$

For $\alpha_G < 1 < \alpha_L$, utility is S-shaped; the risk parameters, $\alpha_G$ and $\alpha_L$, are unbounded below and above, respectively, allowing more flexibility.[35] Marginal utility is bounded at $G = 0$ reaching the values $\lambda R^{\beta-1}$ and $R^{\beta-1}$ just below and above zero. This discontinuous change introduces a true kink in the utility function.

Table 2 provides additional calibrated results using the modified-TK utility specification. The results presented there are not as extreme as those in Table 1 using scaled-TK utility. In particular, the lower marginal utility at small gains has served its purpose of raising the optimal gains threshold, $\Theta$. The estimates in the final row, with $\alpha_G = 0.5$, $\alpha_L = 30$, and $\beta = 0.3$, match the data quite well. As should be obvious from the parameters, this is not an optimized or best fit; rather round numbers are used for $\alpha_G$, $\alpha_L$, and $\beta$ which provide a good fit.

It might well be argued that $\alpha_L = 30$ implies an implausibly high risk-seeking behavior, so the model is doubtful despite fitting the data. Further parameter adjustments cannot do much to improve the fit. For a given threshold-sales policy, $\Theta$-$\theta$, the remaining values in Tables 1 and 2, except the average holding time, are completely determined by the ratio $\mu/\sigma^2$. Therefore, adjusting the utility parameters further cannot better the fit, nor can altering $\mu$ or $\sigma$ improve the fit for *PGR* and *PLR* without degrading that for $Q_G$ and $\varphi_G$. However, the calibration can be improved by introducing additional heterogeneities beyond a difference in the number of stocks held because this model is not one in which a single average investor can serve as a stand-in for the group.

If investors trade some of their stocks differently (*heterogeneous holdings*) or different investors have different sales policies (*heterogeneous investors*), there are further aggregation effects on the various measures. In particular, the closer are $\Theta$ and $\theta$ to one, the shorter will be the average length of each investment episode. Stocks that are traded more frequently will disproportionately affect the statistics because their characteristics will be over-represented. In addition the characteristics of the other stocks held in the same account will also be over-represented as paper gains and losses are counted only when a stock in the same account is sold. The effects of aggregation are described in Propositions 5 and 6. Proposition 5 gives the statistics when different investors follow distinct trading strategies. Proposition 6 gives the statistics when the trading strategies differ for stocks within the same account. These heterogeneities have distinct effects.

**Proposition 5: Realization Utility Statistics with Heterogeneous Investors**. Assume

---

[35] As usual, $\alpha = 0$ corresponds to a logarithmic form, $R^{\beta}\ell n(1+G/R)$. Modified-TK utility can also be adapted to study strictly risk-averse incremental utility by setting $\alpha_L < 1$. Utility is increasing and for $\lambda \geq 1$ strictly concave. If $\alpha_L = \alpha_G$ and $\lambda = 1$, this is incremental power utility; otherwise there is a discontinuous change in risk aversion (if $\alpha_G \neq \alpha_L$) or in marginal utility (if $\lambda \neq 1$) at 0.

**Table 2: Summary Statistics for Reference-Dependent Realization Utility Model with Modified Tversky-Kahneman Utility**

The table reports: $\Theta-1$, $\theta-1$: percentages above and below the reference level for realized gains and losses, $Q_G$: fraction of episodes that end in realized gains, $\varphi_G$: fraction of stocks with unrealized paper gains, $\mathbb{E}[\tau]$: average holding period in trading days, *PGR*, *PLR*: proportions of gains and losses realized, and $\mathcal{O} \equiv PGR/PLR$: Odean's measure. Asset parameters are $\mu = 9\%$ and $\sigma = 30\%$. The accounts' sizes are fixed with $\bar{n} + \sigma_n^2/\bar{n} = 8.0$. Utility parameter are $\lambda = 2$ and $\delta = 5\%$. Transaction costs are $k_s = k_p = 1\%$, and the investor accounts for both costs in his subjective view of his realized gains, $\kappa = K$. Odean's data is taken from Tables 1 and 3 of his 1998 paper. Dhar and Zhu's data is from the notes to their Table 3. The "Fit to Odean's $\Theta$, $\theta$" row uses Odean's estimates of $\Theta$ and $\theta$ to compute the other values using Propositions 2 and 4.

|  | $\Theta-1$ | $\theta-1$ | $Q_G$ | $\varphi_G$ | $\mathbb{E}[\tau]$ | *PGR* | *PLR* | $\mathcal{O}$ |
|---|---|---|---|---|---|---|---|---|
| Odean data | 27.7% | −22.8% | 53.8% | 41.9% | 312 | 14.8% | 9.8% | 1.51 |
| Dhar & Zhu data | ---- | ---- | 65.8% | 46.5% | 122 | 13.2% | 6.4% | 2.06 |
| Fit to Odean's $\Theta$, $\theta$ | 27.7% | −22.8% | 57.7% | 50.7% | 174 | 14.0% | 10.9% | 1.28 |

| Realization Model with Modified-TK Utility |||||||||
|---|---|---|---|---|---|---|---|---|
|  |  | $\Theta-1$ | $\theta-1$ | $Q_G$ | $\varphi_G$ | $\mathbb{E}[\tau]$ | *PGR* | *PLR* | $\mathcal{O}$ |
| $\alpha_G = 0.5$, $\alpha_L = 2.0$ | $\beta = 0$ | 60.4% | −90.7% | 96.3% | 25.2% | 2037 | 35.3% | 0.7% | 50.25 |
|  | $\beta = 0.3$ | 49.2% | never | 100.0% | 17.6% | 2221 | 44.8% | 0 | $\infty$ |
| $\alpha_G = 0.5$, $\alpha_L = 4.0$ | $\beta = 0$ | 44.6% | −64.1% | 85.3% | 31.5% | 909 | 27.9% | 3.0% | 9.36 |
|  | $\beta = 0.3$ | 47.4% | −73.6% | 89.7% | 28.3% | 1169 | 31.1% | 2.0% | 15.45 |
| $\alpha_G = 0.5$, $\alpha_L = 8.0$ | $\beta = 0$ | 27.5% | −42.7% | 77.6% | 33.3% | 351 | 25.0% | 4.6% | 5.44 |
|  | $\beta = 0.3$ | 38.3% | −48.7% | 77.4% | 36.5% | 556 | 23.2% | 4.8% | 4.80 |
| $\alpha_G = 0.5$, $\alpha_L = 30.0$ | $\beta = 0$ | 13.5% | −17.5% | 64.1% | 41.0% | 67 | 18.3% | 8.0% | 2.29 |
|  | $\beta = 0.3$ | 26.7% | −24.3% | 60.5% | 48.0% | 181 | 15.2% | 9.8% | 1.56 |

that asset returns are independent and identically distributed[36] and that investors differ in their trading strategies or number of stocks they hold. Type $i$ investors constitute the fraction $\pi_i$ of the sample, hold $n_i$ stocks, and follow a $\Theta_i$-$\theta_i$ threshold strategy or a $\rho_i$ random strategy. As the number of observed trades increases, the probability limits of the various aggregate statistics are the weighted averages

$$\text{plim}\,\bar{\Xi} = \sum_i w_i \Xi_i \qquad \text{where} \qquad w_i \equiv \frac{\pi_i n_i / \mathbb{E}[\tau_i]}{\sum_i \pi_i n_i / \mathbb{E}[\tau_i]} \qquad (20)$$

---

[36] The assets' means and variances can differ across the types of investors, provided they are identical within types. The effects of any asset differences are completely incorporated into $Q_G^i, Q_L^i, \varphi_G^i, \varphi_L^i$, and $\mathbb{E}[\tau_i]$.

and $\Xi$ is any of the statistics $\Theta$, $\theta$, $Q_G$, $Q_L$, or $\mathbb{E}[\tau]$. The probability limits of the fraction of unrealized paper gains or losses are

$$\text{plim}\,\overline{\varphi}_G = 1 - \text{plim}\,\overline{\varphi}_L = \frac{\sum_i \pi_i n_i (n_i - 1)\varphi_G^i / \mathbb{E}[\tau_i]}{\sum_i \pi_i n_i (n_i - 1) / \mathbb{E}[\tau_i]}. \tag{21}$$

The probability limits of *PGR*, *PLR*, and the Odean measure are

$$\text{plim}\,PGR = \frac{\sum_i \pi_i n_i Q_G^i / \mathbb{E}[\tau_i]}{\sum_i \pi_i n_i [Q_G^i + (n_i - 1)\varphi_G^i] / \mathbb{E}[\tau_i]} \qquad \text{plim}\,PLR = \frac{\sum_i \pi_i n_i Q_L^i / \mathbb{E}[\tau_i]}{\sum_i \pi_i n_i [Q_L^i + (n_i - 1)\varphi_L^i] / \mathbb{E}[\tau_i]}. \tag{22}$$

As before, plim $\mathcal{O}$ = plim *PGR*/plim *PLR*. ∎

**Proposition 6: Realization Utility Statistics with Heterogeneous Holdings**. Assume that a representative investor trades *N* stocks whose returns are independently distributed. These stocks are grouped into categories. Within group *i*, there are $n_i$ stocks with identical means, variances and trading strategies. The latter are either $\Theta_i$-$\theta_i$ threshold strategies or $\rho_i$ random strategies. As the number of observed trades increases, the probability limits of the various aggregate statistics are the weighted averages

$$\text{plim}\,\overline{\Xi} = \sum_i w_i \Xi_i \qquad \text{where} \quad w_i \equiv \frac{n_i / \mathbb{E}[\tau_i]}{\sum_i n_i / \mathbb{E}[\tau_i]} \tag{23}$$

and $\Xi$ is any of the statistics $\Theta$, $\theta$, $Q_G$, $Q_L$, or $\mathbb{E}[\tau]$. The probability limits of the fraction of unrealized paper gains or losses are

$$\text{plim}\,\overline{\varphi}_G = 1 - \text{plim}\,\overline{\varphi}_L = \frac{\sum_i n_i (\sum_j n_j \varphi_G^j - \varphi_G^i) / \mathbb{E}[\tau_i]}{(N-1)\sum_i n_i / \mathbb{E}[\tau_i]}. \tag{24}$$

The probability limits of *PGR*, *PLR*, and the Odean measure are

$$\text{plim}\,PGR = \frac{\sum_i n_i Q_G^i / \mathbb{E}[\tau_i]}{\sum_i n_i [Q_G^i + \sum_j n_j \varphi_G^j - \varphi_G^i] / \mathbb{E}[\tau_i]} \quad \text{plim}\,PLR = \frac{\sum_i n_i Q_L^i / \mathbb{E}[\tau_i]}{\sum_i n_i [Q_L^i + \sum_j n_j \varphi_L^j - \varphi_L^i] / \mathbb{E}[\tau_i]} \tag{25}$$

with plim $\mathcal{O}$ = plim *PGR*/plim *PLR* as always. ∎

Table 3 summarizes the calibrated results for heterogeneous investors and heterogeneous holdings based on the statistics derived in Proposition 5 and 6. For comparison purpose, we pick the utility parameters used in Table 2, excluding the high risk tolerance case which is no longer needed for a good fit. All of the averages in this table assume there is an equal mix of two types. For "heterogeneous investors," one half of the investors optimize realization utility and the other half trade randomly. For "heterogeneous holdings," each investor trades one-half of his stocks by optimizing his realization utility and trades the other half randomly.

**Table 3: Summary Statistics for Mixed Reference-Dependent Realization Utility and Random Trading**

The table reports: $\Theta-1$, $\theta-1$: percentages above and below the reference level for realized gains and losses, $Q_G$: fraction of episodes that end in realized gains, $\mathbb{E}[\tau]$: average holding period in trading days, $\varphi_G$: fraction of stocks with unrealized paper gains, *PGR*, *PLR*: proportions of gains and losses realized, and $\mathcal{O} \equiv PGR/PLR$: Odean's measure. Asset parameters are $\mu = 9\%$ and $\sigma = 30\%$. Transaction costs are $k_s = k_p = 1\%$ with $\kappa = K$. Odean's data is taken from Tables 1 and 3 of his 1998 paper. Dhar and Zhu's data is from the notes to their Table 3. The "Fit to Odean's $\Theta$, $\theta$" row uses Odean's estimates of $\Theta$ and $\theta$ to compute the other values using Propositions 2 and 4. The accounts' sizes are fixed at 8 which matches $\bar{n} + \sigma_n^2/\bar{n}$ in the previous two tables. In each row under Heterogeneous Investors, one half of the investors are realization-utility traders with utility parameters $\lambda = 2$, $\delta = 5\%$, and others as stated. The other half of the investors trade randomly with the stated Poisson intensity. In each row under Heterogeneous Holdings, each investor trades half his stocks to optimize realization utility and the other half randomly.

| | | | $\bar{\Theta}-1$ | $\bar{\theta}-1$ | $\bar{Q}_G$ | $\mathbb{E}[\tau]$ | Heterogeneous Investors $\bar{\varphi}_G$ | PGR | PLR | $\mathcal{O}$ | Heterogeneous Holdings $\bar{\varphi}_G$ | PGR | PLR | $\mathcal{O}$ |
|---|---|---|---|---|---|---|---|---|---|---|---|---|---|---|
| | Odean data | | 27.7% | −22.8% | 53.8% | 312 | 41.9% | 14.8% | 9.8% | 1.51 | 41.9% | 14.8% | 9.8% | 1.51 |
| | Dhar & Zhu data | | ---- | ---- | 65.8% | 122 | 46.5% | 13.2% | 6.4% | 2.06 | 46.5% | 13.2% | 6.4% | 2.06 |
| | Fit to Odean's $\Theta$, $\theta$ | | 27.7% | −22.8% | 57.7% | 174 | 50.7% | 14.0% | 10.9% | 1.28 | 50.7% | 14.0% | 10.9% | 1.28 |
| $\alpha_G = 0.5$, $\alpha_L = 2.0$ | $\beta=0$ | $\rho=1.5$ | 26.1% | −19.5% | 57.5% | 308 | 52.1% | 13.6% | 11.3% | 1.21 | 38.0% | 17.8% | 8.9% | 1.99 |
| | | $\rho=1$ | 34.1% | −24.2% | 59.8% | 445 | 52.0% | 14.1% | 10.7% | 1.32 | 38.5% | 18.1% | 8.6% | 2.12 |
| | $\beta=0.3$ | $\rho=1.5$ | 25.1% | −19.7% | 57.5% | 310 | 51.7% | 13.7% | 11.2% | 1.23 | 33.7% | 19.6% | 8.4% | 2.34 |
| | | $\rho=1$ | 32.7% | −24.5% | 59.8% | 449 | 51.5% | 14.2% | 10.6% | 1.35 | 34.3% | 19.9% | 8.0% | 2.48 |
| $\alpha_G = 0.5$, $\alpha_L = 4.0$ | $\beta=0$ | $\rho=1.5$ | 26.6% | −21.5% | 59.1% | 282 | 50.8% | 14.3% | 10.6% | 1.34 | 41.8% | 16.8% | 9.1% | 1.84 |
| | | $\rho=1$ | 33.8% | −26.4% | 61.7% | 392 | 50.1% | 15.0% | 9.9% | 1.51 | 42.4% | 17.2% | 8.7% | 1.99 |
| | $\beta=0.3$ | $\rho=1.5$ | 26.3% | −21.2% | 58.7% | 292 | 51.1% | 14.1% | 10.8% | 1.31 | 39.9% | 17.4% | 8.9% | 1.94 |
| | | $\rho=1$ | 33.8% | −26.2% | 61.3% | 412 | 50.5% | 14.8% | 10.0% | 1.47 | 40.6% | 17.8% | 8.5% | 2.09 |
| $\alpha_G = 0.5$, $\alpha_L = 8.0$ | $\beta=0$ | $\rho=1.5$ | 24.6% | −23.0% | 61.8% | 226 | 47.6% | 15.7% | 9.4% | 1.66 | 43.3% | 16.9% | 8.8% | 1.93 |
| | | $\rho=1$ | 29.4% | −27.1% | 64.5% | 292 | 46.1% | 16.7% | 8.6% | 1.94 | 44.0% | 17.3% | 8.3% | 2.09 |
| | $\beta=0.3$ | $\rho=1.5$ | 26.7% | −21.8% | 59.6% | 256 | 50.2% | 14.5% | 10.4% | 1.40 | 44.7% | 16.0% | 9.4% | 1.69 |
| | | $\rho=1$ | 33.2% | −26.1% | 62.1% | 345 | 49.5% | 15.2% | 9.7% | 1.57 | 45.4% | 16.4% | 9.0% | 1.81 |

Throughout our analysis, stock-level narrow framing is assumed. That is, decisions on when to sell are not affected by any other stock's performance. Therefore, all empirical statistics for *realized* gains and losses depend only on the overall distribution of investing strategies and stock parameters; the form of the heterogeneity, whether it is within or across accounts, is irrelevant. Specifically, if the stock-level heterogeneity is the same in the "heterogeneous investors" and the "heterogeneous holdings" cases, then $\overline{\Theta}, \overline{\theta}, \overline{Q}_G, \overline{Q}_L$ and $\mathbb{E}[\tau]$ are identical. However, in Odean's methodology, *paper* gains and losses are counted by the outside econometrician only when another stock in the same account is sold. As a result, the difference between "heterogeneous investors" and "heterogeneous holdings" in grouping stocks into accounts yields distinct values for the statistics related to paper gains and losses, i.e., $\overline{\varphi}_G, \overline{\varphi}_L$, *PGR*, *PLR*, and $\mathcal{O}$.

Following Odean, $\overline{\varphi}_G$ ($\overline{\varphi}_L$) is the fraction of stocks trading at a paper gain (loss) when some other stock in the same account is sold. Stocks with a lower $\mathbb{E}[\tau]$ trade more frequently and therefore increase the impact of all other stocks *held in the same account* on $\overline{\varphi}_G$ and $\overline{\varphi}_L$ and the other statistics that depend on these. For example from Table 2, investors with $\alpha_L = 8$, $\alpha_G = 0.5$, and $\beta = 0$ have an average holding period of 351 trading days with $\varphi_G = 33.3\%$; $\rho = 1$ random trades have a smaller average holding period of 250 days and a higher $\varphi_G$ of 55.3%. Therefore, with an equal mixture of these investors, the average $\overline{\varphi}_G = 46.1\%$ is closer to that of the random traders since they trade more often. Conversely with heterogeneous holdings, the opposite is true; the average $\overline{\varphi}_G = 44.0\%$ is closer to the threshold-traded stock value. Since random trades occur more often, the threshold-traded stocks are the ones observed more often in determining the paper gains and losses. As a consequence *PGR* and $\mathcal{O}$ are typically larger and *PLR* is typically smaller for heterogeneous holdings than for heterogeneous investors.

The reported averages are in fairly close agreement with Odean's empirical results for each level of risk tolerance with no need to resort to the very high risk tolerance required in Table 2. The average time between trades has also increased to better match Odean's value. This table simply highlights the possibilities. Using mixtures of other than 50-50, including more types of investors, or allowing heterogeneity of trading strategies both across investors and within the same account would permit further tweaking of the fit.

In summary, we have shown that our realization utility model is consistent with and can shed light on several dimensions of observed trading data. We have made no attempt to match the empirical patterns exactly, nor have we considered all dimensions. For instance, in his 1998 paper, Odean reports the average size of paper gains and losses measured when some other stock in the same account was traded. The average paper gain and loss were 46.6% and –39.3%, which are larger in magnitude than the realized gains and losses of 27.7% and –22.8%. Our model with a single representative investor following an identical $\Theta$-$\theta$ strategy on all the stocks cannot possibly generate this pattern as the realizations occur at the extreme points of the paper gains and losses distribution. Although it is possible to generate this pattern by considering an economy with heterogeneous investors, other explanations are also possible such as investors updating their reference levels based on the recent price history of the asset.[37] Once the reference level changes from the initial purchase price, the subjectively measured gain or loss realized by an investor differs from the gain or loss as measured objectively by an outside econometrician. This topic is examined in detail in Ingersoll and Jin (2012).

---

[37] See section IV of Odean (1999) for an informal discussion on this. Related experimental and empirical evidence can be found in Gneezy (2005) and Arkes, Hirshleifer, Jiang, and Lim (2008, 2010).

**4. Voluntary Loss Realization**

One of the main objectives of this paper is to create an intertemporal realization utility model with voluntary loss realization. In earlier sections we introduced scaled- and modified-TK utility functions, and our numerical analysis shows that either can generate voluntary loss taking. In this section, we pose a more general question: what are the necessary characteristics for *any* utility function to generate voluntary sales at losses in the intertemporal realization utility framework posed in (3)? We are not ruling out other possible preference or belief-based explanations that contribute to voluntary loss taking such as changes in information or portfolio rebalancing. Nevertheless, our general analysis may shed light on theoretical and experimental work on realization utility and the disposition effect.

Both versions of TK utility have two properties that separately seem important, an S-shape and reference scaling. To illustrate, consider an investor with the scale-free utility function, $U(G) = \text{sgn}(G) \cdot |G|^{1/2}$ that does not explicitly depend on the reference level. This investor might be willing to realize a loss of 2 since, if the price subsequently recovers, he can take two gains of 1 and have positive total utility. Not taking the loss of 2 prevents realizing the recovery as a gain. With scale-free concave utility, the benefit of a recovery gain can never offset the disutility of the loss because marginal utility is decreasing so the disutility of the loss must be larger than the utility of subsequent gains no matter how it is divided. Now consider a simple scaled utility function with no convexity for losses, $U(G, R) = G/R$. A loss from 4 to 2 has a disutility of $-2/4$ while a recovery from 2 back to 4 provides a utility of $2/2$. Now the decreasing scaling may make taking the loss worthwhile in order to realize a later gain. Conversely, with increasing scaling, loss taking could never be optimal. This analysis is made precise in Proposition 7 below which describes the conditions under which losses will never be realized voluntarily.

**Proposition 7:** Assume that an investor maximizes expected realization utility as in (3) with a burst utility function of $U(G, R)$ and that the reference level is updated to the investment value after a sale but remains constant between sales. The following four conditions

$$\text{(i) } \delta > 0 \quad \text{(ii) } \frac{\partial U}{\partial G} > 0 \quad \text{(iii) } \frac{\partial^2 U}{\partial G^2} \leq 0 \quad \text{(iv) } \frac{\partial^2 U}{\partial G \, \partial R} \geq 0 \quad (26)$$

are jointly sufficient to preclude the voluntary realization of any losses in the absence of transaction costs.[38] ∎

---

[38] As seen in Figure 5, transaction costs widen the no-sales region and therefore make voluntary loss taking less likely to occur. With transaction costs we would require stronger violations of these conditions to make voluntary loss taking optimal.

A proof of this theorem is given in the Appendix. The intuition for the result is that the disutility of a loss cannot be offset by the utility of any later gains that recoup this loss because each gain utility burst comes at a later time, has a smaller marginal utility, and has a lower reference level. Each of these aspects makes the gain utility smaller by properties (i), (iii), and (iv), respectively.

The first three conditions in (26) are standard utility properties. Time preference is positive, and marginal utility is positive and decreasing. Taking the first two as inviolable, voluntary losses are then possible only if utility is not everywhere concave or marginal utility is not everywhere increasing in the reference level. We have already discussed the option-like effect that S-shaped utility has on loss taking from resetting the reference level. A violation of condition (iv), that is, marginal utility is decreasing in the reference level over some range, can have the same effect. A loss measured from a high $R$ can have a smaller negative impact on utility than the same size gain from a lower $R$. This might induce the investor to realize a loss so he is in position to take a later gain when the asset recovers in value.

S-shaped utility functions are commonly used in behavioral models, and we have already argued that $\partial |U(G, R)|/\partial R \leq 0$ is a likely description of realization utility which, if true, means (iv) cannot hold except when $\partial^2 U/\partial G \partial R \equiv 0$.[39] Therefore, both S-shaped utility and decreasing scaling of utility with respect to $R$ may contribute to the optimal voluntary realization of losses. In theory a violation of (iii) or (iv) alone is sufficient to make voluntary loss taking optimal. However, our model calibration indicates that both are probably necessary to explain the data.

## 5. Further Model Predictions and Applications

As discussed above, our paper makes several direct and specific predictions about trading activity.[40] The two-point, $\Theta$-$\theta$, trading strategy is quite specific about the volume of gains and losses, the holding periods, and the sizes of both realized and paper gains and losses. These results can help to explain the difference between the trading volume in rising and falling markets as well as the effect of historical highs on the propensity to sell. Together, risk-seeking behavior and the trading strategy might also explain the heavy trading of highly valued assets since the optimal strategies are related to the assets' means and variances. Furthermore, if an investor's subjective reference level is not constant but is updated based on recent stock prices, then the predicted trading patterns become path-dependent.

Models like this one may also rationalize hazard-rate types of models for investor behavior. In a recent paper, Ben-David and Hirshleifer (2012) document a V-shape empirical pattern between the probability of selling a stock and the unrealized return since purchase for fixed holding periods. Using the statistics developed in Proposition 5 for heterogeneous investors, our model can match this pattern.[41] Many of these considerations are examined further in Ingersoll and Jin (2012).

---

[39] Letting subscripts denote partial derivatives, $U_2(G, R) = U_2(0, R) + \int_0^G U_{12}(g, R)dg$. Since $U(0, R) = 0$ for all $R$, $U_2(0, R) = 0$. If (iv) holds strictly, i.e., $U_{12} > 0$, then the integral is positive (negative) for positive (negative) $G$, and $\partial |U|/\partial R > 0$. Therefore, assuming (iv), the relation $\partial |U|/\partial R \leq 0$ cannot hold except when $U_{12} \equiv 0$.

[40] A more detailed discussion on some of these applications can be found in Barberis and Xiong (2012).

Tax-trading behavior is another obvious topic addressed by this model. In fact, capital gains are a near perfect fit with realization utility since they are typically due only upon sale of an asset. Standard reasoning indicates that an investor should realize losses and defer gains to minimize the tax burden. With taxes loss taking should predominate over gain taking which is the opposite of the disposition effect. However, this reasoning again fails to recognize the importance of the reinvestment effect. To illustrate, consider an investor with modified-TK utility with $\alpha_G = 0.5$, $\alpha_L = 4$, $\beta = 0.3$, $\lambda = 2$, $\delta = 5\%$ investing in an asset with $\mu = 9\%$ and $\sigma = 30\%$ and paying 1% transaction costs. In the absence of a capital gains tax, the investor would sell at $\Theta = 1.460$ or $\theta = 0.261$.[42] With a 15% capital gains tax, he sells at $\Theta = 1.549$ or $\theta = 0.248$. The capital gains tax does cause the investor to postpone the realization of gains because they now provide a smaller utility burst. However, loss taking also is postponed precisely because gains are less valuable so it no longer is quite so advantageous to realize a painful loss to reset the reference point. Of course, loss taking is affected less than gain taking because of different tax treatment.

Our model also makes other indirect predictions. Two such predictions are the flattening of the capital market line and the pricing of idiosyncratic risk. There is no equilibrium model in our paper so precise predictions are not possible, but the model does indicate directional effects.

The $\mu$-$\sigma$ indifference curves for a realization-utility investor are flatter than the observed capital market line and, in some cases, are actually decreasing with the investor preferring more variance to less. But out model does not address to what these indifference curves should be applied. If an investor holds only a diversified mutual fund, the model indicates he should display less risk aversion than typical in the selection of a fund. But it also seems plausible that investors might save the bulk of their wealth in diversified portfolios recognizing that those are the best vehicle for long-term saving[43] and still actively trade other stocks because they enjoy doing so. Only this latter investment activity might be governed by realization utility of the type we have modeled. This might explain why we see only a small number of stocks in typical trading accounts.

With flatter indifference curves, there will be an excess demand for high-variance stocks. This means that high-beta and high residual risk stocks should have smaller expected returns than predicted by equilibrium models like the CAPM. Ang, Hodrick, Xing, and Zhang (2006) document this. They test alphas for stocks with different total and residual volatilities and find just this result. The difference between the alphas of the highest and lowest volatility stocks is −1.35% ($t = -4.62$) for the CAPM and −1.19% ($t = -5.92$) for Fama-French three-factor model. The difference in alphas between the highest and lowest residual volatility stocks is −1.38% ($t = -4.56$) for the CAPM and −1.31 ($t = -7.00$) for FF-3.

---

[41] Ben-David and Hirshleifer (2012) implicitly assume that investors trading is random with hazard rate (intensity $\rho$) that is a function of holding period and the size of the unrealized gain. In contrast in our model with heterogeneous investors, the representative investor is not an average investor, and the hazard rate measures an aggregation effect.

[42] For consistency in the comparison, we assume that the investor views his subjective gain as the taxable gain and resets his reference level to the new cost basis even if the tax rate is zero. That is, he uses the gross cost view of setting his reference level as discussed in footnote 8.

[43] In a Swedish data set, Calvet, Campbell, and Sodini (2009) find that the asymmetry between selling winners and selling losers is much weaker for investors' holdings of mutual funds than for their individual stocks, suggesting that households' motive for mutual funds investment is different from individual stocks investment.

Precise predictions are, unfortunately, not possible, because we have ignored modeling diversification directly. In a standard portfolio model, risk-averse investors optimally hold many assets to provide diversification. Our model assumes narrow framing with the utility from each investment depending solely on how it is traded. It might seem, therefore, that there is no benefit from diversification. However, this is not true. For utility that is homogeneous of degree β as we have assumed, the value of investing wealth $W$ in several assets in proportions $w_i$ and subsequently reinvesting each portion without any rebalancing is $\sum_i (w_i W)^\beta v_i(1)$ where $v_i$ is the separate valuation of asset $i$ that depends on its mean and variance and the specific utility function. For $0 < \beta < 1$, utility is maximized with $w_i \propto [v_i(1)]^{1/(\beta-1)}$. Of course, this portfolio is generally not optimal absent the just-assumed no-rebalancing restriction, but it does demonstrate that investing in a single stock is a dominated policy so some diversification must be optimal. A more thorough analysis would show that when rebalancing is allowed, investors should deviate from a strictly constant Θ-θ policy whenever their allocations stray too far from the optimal.

However, before diversification and rebalancing can be studied completely, a more fundamental question must be answered: How are separate gains and losses aggregated under realization utility? Our model assumes complete narrow framing both across assets and over time. All utility bursts are evaluated separately and then their discounted values are summed over different assets and time periods. In a different model investors might aggregate contemporaneous gains and losses into a single utility burst and sum those over time. Both assumptions are psychologically plausible, but either creates problems in multi-period models, particularly continuous-time models.

If the utility function aggregates contemporaneous gains and losses, and the investor's utility is S-shaped, losses on different assets should be taken simultaneously if possible because marginal utility is decreasing in the magnitude of losses. This will lead to timing complications with investors postponing some loss taking while accelerating other losses to achieve synchronicity. On the other hand, gains on different assets should always be kept separate in time since their marginal utility is decreasing. This is trivial to achieve in continuous time with time-additive utility which so narrowly frames the time dimension that a separation of $dt$ is sufficient for a completely independent evaluation. But is it reasonable to assume that two gains realized simultaneously have a different utility than the same two gains realized only instants apart? And if not, how long a time separation is required? Similar questions have arisen in the standard consumption-portfolio problem and various suggestions like recursive utility or Hindy, Huang, and Kreps (1992) intertemporal aggregation have been made. The question here is more difficult because consumption is naturally smoothed in intertemporal models, but here we have both smoothing of gains and lumping of losses to address.

# 6. Concluding Remarks

In this paper, we have built an intertemporal realization utility model to study investors' trading behavior. Highlighting the role reinvestment plays in a dynamic context, we have shown that investors may voluntarily realize losses even though this has an immediate negative utility impact. The necessary condition for voluntary losses is either risk seeking behavior over at least some losses or decreasing scaling under which the magnitude of the utility of gains and losses realized with smaller reference levels is larger.

Under our model a two-point sales strategy is optimal; an investor sells for a gain when the asset value rises to a fixed multiple of the reference level or sells at a loss at a fixed fraction of the reference level. We have provided a detailed calibration of the model showing that the trading data of Odean and others is in close agreement with such two-point strategies but is inconsistent with random trading that is independent of the potential gain. We also introduced a modified form of Tversky-Kahneman utility which predicts, either alone or in a model of heterogeneous investors, the average realized gains and losses observed in the data.

We have discussed some properties predicting trading patterns and price effects. In particular, our model suggests a flattening of the capital market line and that idiosyncratic risk could have a negative risk premium. Both of these features also are seen in the data. But other effects still need to be investigated. For example, how do trading patterns and volume evolve over time in different markets? What is the relation between realization utility and momentum?

There are several directions for future theoretical research. One important step is to study the diversification motive of realization-utility investors and solve a full portfolio problem. We do know that some diversification and rebalancing is optimal, but this means that the optimal sales strategy for a stock depends not only on its reference level but also on the prices and reference levels of the other assets held in the portfolio. In addition, assumptions must then be made on how contemporaneous and near-contemporaneous gains and losses are subjectively aggregated and about the proper reference level for a position that consists of shares purchased for different prices at different times.

It is also important to understand how realization utility interacts with other types of utility. Do investors also receive utility just from holding assets with paper gains even if they do not sell? Clearly investors also value consumption; are the motives to smooth consumption and to realize gains evaluated separately or combined somehow?

Finally if an S-shaped utility function is important, does probability weighting also have an effect on realization utility? It is a non-trivial task to incorporate probability weighting into an intertemporal setting because the law of iterated expectations does not hold if probabilities are replaced by decision weights. As shown in Barberis (2012) and Ingersoll (2012), cumulative probability weighting typically induces time inconsistency, and certain rules that define economic actions need to be imposed in order to further model this type of behavior.

# Appendices

## A: Notation

<div style="text-align: center">TABLE OF IMPORTANT NOTATION</div>

===========================================================================

| | |
|---|---|
| $\mathbb{E}$ | Expectations operator |
| $G$ | Dollar size of a gain or a loss |
| $k_s$ | Proportional transaction cost for selling stock |
| $k_p$ | Proportional transaction cost for purchasing stock |
| $K$ | Round-trip proportional transaction cost for selling the current asset being held and reinvesting in another asset, $K \equiv (1 - k_s)/(1 + k_p)$ |
| $\bar{n}$ | Average number of stocks held in a single investment account |
| $\mathcal{O}$ | The Odean measure of the disposition effect, $\mathcal{O} \equiv PGR/PLR$ |
| $PGR$ | Proportion of gains realized |
| $PLR$ | Proportion of losses realized |
| $R$ | The reference level |
| $Q_G$ | The probability that a given investment episode eventually ends with a realized gain |
| $Q_L$ | The probability that a given investment episode eventually ends with a realized loss |
| $S$ | The stock price |
| $U$ | The utility burst function $U(G, R)$ |
| $u$ | The reduced form utility burst function $u(G/R) \equiv R^{-\beta} U(G, R)$ |
| $V$ | The value function $V(X, R)$ |
| $v$ | The reduced form value function $v(x) \equiv R^{-\beta} V(X, R)$ |
| $X$ | The level of investment |
| $x$ | Gross return per dollar of the reference level, $x \equiv X/R$ |
| $\alpha_G$ | Parameter of risk aversion for gains in evaluating utility bursts |
| $\alpha_L$ | Parameter of risk seeking for losses in evaluating utility bursts |
| $\beta$ | The scaling parameter gauging the impact of the reference level, $R$, on utility bursts |
| $\gamma_1, \gamma_2$ | Characteristic roots of the partial differential equation, defined in equation (8) |
| $\delta$ | The subjective discount rate |
| $\eta$ | Defined parameter $\eta \equiv 1 - 2\mu/\sigma^2$ |
| $\Theta$ | Optimal sales multiple for a gain; sale occurs at $X = \Theta R$ |
| $\theta$ | Optimal sales fraction for a loss; sale occurs at $X = \theta R$ |
| $\kappa$ | Parameter measuring the subjective effect of transaction costs in evaluating utility bursts |
| $\lambda$ | Loss aversion parameter |
| $\mu$ | Growth rate of the stock price |
| $\rho$ | Poisson intensity for random trading |
| $\sigma$ | Logarithmic standard deviation of the stock price |
| $\sigma_n$ | Standard deviation of number of stocks held in investment accounts |
| $\tau$ | Duration of an investment episode from purchase to sale |
| $\varphi_G$ | The fraction of time an asset has an unrealized paper gain |
| $\varphi_L$ | The fraction of time an asset has an unrealized paper loss |

## B: Verification of the Optimality of Constant Proportional Sales Policies

Consider the general sale and reinvestment problem given in the text. For any given realized stochastic price path and sales policy denote the original stock price by $S_0$ and the stock prices at the points of sale, occurring at times $t_1, t_2, \ldots, t_n$, by $S_1, S_2, \ldots, S_n$, etc. The number of shares purchased at price $S_n$ and sold at price $S_{n+1}$ is $N_n$. The dollar amount under investment is $X_n^+ = N_n S_n$ just after the purchase at $t_n$ and $X_n^- = N_{n-1} S_n$ just before the sale that finances this. These amounts differ by the round trip transaction cost so the relations among the $X$'s are given by the recurrences

$$X_n^+ = (1-k_s)X_n^-/(1+k_p) \equiv K X_n^- \qquad X_{n+1}^- = X_n^+ S_{n+1}/S_n. \tag{A1}$$

The reference level, $R_n$, established after the sale and repurchase at $t_n$ is $X_n^+$ if the investor considers both sets of transaction costs. Alternatively he might ignore the repurchase costs setting $R_n = X_n^-(1-k_s) = X_n^+(1+k_p)$. To cover both of these and many other cases we define $R_n = K'X_n^+$. The subjective rate of return realized on the gain at time $t_n$ and its utility burst are

$$\begin{aligned}\kappa X_n^-/R_{n-1} - 1 &= (\kappa/K')X_n^-/X_{n-1}^+ - 1 = (\kappa/K')S_n/S_{n-1} - 1\\ & e^{-\delta t_n} R_{n-1}^\beta u\left((\kappa/K')S_n/S_{n-1} - 1\right).\end{aligned} \tag{A2}$$

Using the recursion relations in (A1)

$$R_n = K'X_n^+ = K'KX_n^- = K'KX_{n-1}^+ S_n/S_{n-1} = \cdots = K'K^n X_0^+ S_n/S_0. \tag{A3}$$

Total utility is therefore

$$\Upsilon = (X_0^+ K')^\beta \sum_{j=1} e^{-\delta t_j}[K^{j-1}(S_j/S_0)]^\beta u\left((\kappa/K')S_j/S_{j-1} - 1\right). \tag{A4}$$

The same relation holds looking ahead from any point in the future.

Now consider a rule that generates the optimal sales policy. It sets gain and loss sales points for the first sale of $\overline{S}_1(S_0)$ and $\underline{S}_1(S_0)$ and sets contingent rules for the second sales points, $\overline{S}_2(S_1)$ and $\underline{S}_2(S_1)$, etc. These optimal policies cannot depend on time since the stochastic process is time homogeneous and the investor has a constant subjective discount rate. Nor can the optimal policies depend on the current state, $X_0^+$, since from (A4) realized utility depends only on the proportional scaling factor $(X_0^+)^\beta$ and the price ratios which have stochastic constant returns to scale and are independent. Consequently, the optimal policy must be a constant sales policy.

It is clear from this analysis that constant proportional transaction costs, a constant and proportional subjective interpretation of the realized return and reference level ($\kappa$ and $K'$), a stock price process with independent and identically distributed returns, a constant rate of time preference, an infinite horizon, and a utility based on rates of return with power scaling factor are all necessary for a constant optimal policy.

## C: Transversality Conditions

There are several different conditions required in our analysis to keep the value function finite and produce a well-defined optimal trading strategy. First the discount rate must be large enough or utility bursts far in the future will dominate the value function and make it unbounded. This restriction mimics the transversality conditions in the standard portfolio problem. Second the scaling parameter β must be not too large otherwise repeated selling lets total utility accumulate too quickly by increasing the reference level and thereby the utility bursts of future sales. Finally, transaction costs cannot be zero or the investor can repeatedly realize small gains with their very high marginal utility.

To see that δ and β must be restricted, suppose an investor adopts a constant sales policy, Θ, and never sells at a loss. With each sale the reference level increases by the factor $K\Theta$ so the reference level for $n^{th}$ sale will be $R_n = R_1(K\Theta)^{n-1}$. The $n^{th}$ sale has a subjective gain of $\kappa\Theta - 1$ per dollar of the reference level. The expected lifetime utility from a series of sales at gains and no sales at losses is

$$\mathbb{E}\sum_{n=1}^{\infty} e^{-\delta \tilde{t}_n} R_n^\beta u(\kappa\Theta - 1) = u(\kappa\Theta - 1)\frac{R_1^\beta}{(K\Theta)^\beta}\sum_{n=1}^{\infty}(K\Theta)^{n\beta}\mathbb{E}[e^{-\delta \tilde{t}_n}]. \tag{A5}$$

Here $\tilde{t}_n$ is the random time of the $n^{th}$ sale. Note that the utility bursts realized are known; it is only the timing that is random. Since the times between successive sales are independent and identically distributed

$$\mathbb{E}[e^{-\delta \tilde{t}_n}] = \mathbb{E}[e^{-\delta \tilde{t}_1}]\mathbb{E}[e^{-\delta(\tilde{t}_2-\tilde{t}_1)}]\cdots\mathbb{E}[e^{-\delta(\tilde{t}_n-\tilde{t}_{n-1})}] = \left(\mathbb{E}[e^{-\delta \tilde{t}_1}]\right)^n, \tag{A6}$$

and the final sum in (A5) is

$$\sum_{n=1}^{\infty}\left((K\Theta)^\beta \mathbb{E}[e^{-\delta \tilde{t}_1}]\right)^n \tag{A7}$$

which converges if and only if $(K\Theta)^\beta \mathbb{E}[e^{-\delta \tilde{t}_1}] < 1$. The expected value in (A7) depends on Θ through the stopping time $\tilde{t}_1$, but unless δ > 0, it is at least one for any Θ so the sum is unbounded for any Θ > 1/K and infinite utility is possible. Therefore δ must be positive just as in the standard infinite-horizon investment problem.

The expectation in (A7) is a particular value of the Laplace transform[44] of the first-passage time density of the random variable, $X_t$, to Θ$R$; that is,

---

[44] The Laplace transform of a density function, $\mathcal{L}(f(t)) \equiv \mathbb{E}[e^{-st}]$, can be easily determined from the moment generating function, $\mathcal{M}(f(t)) \equiv \mathbb{E}[e^{st}]$, for a negative argument. The Laplace transform is defined for all values of μ (unlike the moment generating function) since the first passage time is a positive random variable.

$$\mathbb{E}[e^{-s\tilde{t}_1}] = \exp\left(\sigma^{-2}\ell n\Theta\left[\mu - \tfrac{1}{2}\sigma^2 - \sqrt{(\mu - \tfrac{1}{2}\sigma^2)^2 + 2s\sigma^2}\right]\right) \quad (A8)$$
$$\Rightarrow \quad \mathbb{E}[e^{-\delta\tilde{t}_1}] = \exp(-\gamma_1 \ell n\Theta)$$

where $\gamma_1$ is the positive exponent in the solution of the differential equation for valuation as given in (8). Therefore, for the sum in (A7) to be bounded, we also require

$$1 > (K\Theta)^\beta \mathbb{E}[e^{-\delta\tilde{t}_1}] = K^\beta \Theta^{\beta-\gamma_1} . \quad (A9)$$

The feasible policy choices are $\Theta > 1/\kappa \geq 1$, so infinite utility can be achieved if $\beta > \gamma_1$ with any feasible $\Theta > K^{-\beta/(\beta-\gamma_1)}$. If $0 \leq \beta \leq \gamma_1$, then the sum converges for all feasible policies.[45] Combining these results, necessary conditions for there to be no investment plans which lead to infinite expected utility are

$$\beta \leq \gamma_1 \quad \text{and} \quad \delta > 0 . \quad (A10)$$

These conditions depend only on the scaling and the discounting of the burst utility function and are required whatever the functional form of the reduced utility $u(\cdot)$.

However, these conditions are not sufficient. Unbounded utility can also be achieved if the growth rate of the asset is too large. Suppose an investor adopts a gains-only policy ($\theta = 0$), then from (10) in the text, $C_2 = 0$. And as $\Theta \to \infty$, the initial value function is

$$v(1) = C_1 = \frac{u(\kappa\Theta - 1)}{\Theta^{\gamma_1} - (K\Theta)^\beta} \sim \begin{cases} \kappa^{\alpha_G}\Theta^{\alpha_G - \gamma_1} & \text{for scaled-TK} \\ \kappa^{\alpha_G}\Theta^{\alpha_G - \gamma_1}/\alpha_G & \text{for modified-TK} \end{cases} \quad (A11)$$

since $\beta \leq \gamma_1$. This is clearly unbounded if $\alpha_G > \gamma_1$; therefore, a necessary condition for no transversality violation is $\alpha_G \leq \gamma_1$ or in terms of the exogenous parameters

$$\alpha_G[\mu + \tfrac{1}{2}(\alpha_G - 1)\sigma^2] \leq \delta . \quad (A12)$$

This restriction is similar to the transversality condition found in the standard infinite horizon portfolio problem. For linear utility over gains, $\mu$ cannot exceed the discount rate, but risk aversion expands the set of admissible values for $\mu$.[46]

---

[45] We have assumed that $\beta \geq 0$ to ensure participation; however, if $\beta < 0$, then the sum in (A7) might diverge for $\Theta \approx 1$. The minimum feasible value for $\Theta$ is $1/\kappa$, so the convergence condition is $1 > K^\beta \kappa^{\gamma_1 - \beta}$. If the investor is subjectively fully cognizant of all transaction costs then $\kappa = K$, and this condition is met; otherwise for $\kappa > K$, the lower bound on $\beta$ is $\gamma_1 \ell n(\kappa)/\ell n(\kappa/K)$.

[46] The interpretation of this transversality violation is that for any sufficiently high policy, $\Theta$, choosing an even higher policy will increase expected utility. As $\Theta \to \infty$, sales will become increasingly rare but never cease altogether so expected utility continues to rise. This limiting result should not be confused with the *ex ante* policy of never selling, which is *assigned* zero utility.

Finally unbounded utility can be achieved for scaled-TK utility[47] if there are no transaction costs so there is also no well-defined optimal strategy in the absence of costs. To demonstrate this, we construct a sequence of feasible sales strategies and show that as the transaction costs go to zero, the constructed strategies lead to unbounded utility. Therefore, the optimal strategy must also yield infinite utility in the limit of zero costs.

Consider a sequence of economies indexed by $k$. We assume that the sales and purchase costs are in the same proportion for each step of the sequence, that is, $k_s = k$ and $k_p = ck$ for some $c \geq 0$. At each step the sales policies considered are

$$\Theta = 1/(1-k) + k^{\omega_0}, \qquad \theta = 1 - k^{\omega_1}, \quad \text{where } 0 < \omega_0, \omega_1 < \tfrac{1}{2}. \tag{A13}$$

These are not assumed to be the optimal strategies only feasible ones.

Initial utility is $v(1) = C_1 + C_2$ where these constants are defined in (10). Using a Taylor expansion for small $k$, the functions determining $C_1$ and $C_2$ are[48]

$$\begin{aligned}
c_1(\Theta) &= (\gamma_1 - \beta)k^{\omega_0} + \tfrac{1}{2}[\gamma_1(\gamma_1 - 1) - \beta(\beta - 1)]k^{2\omega_0} + o(k^{2\omega_0}) \\
c_1(\theta) &= (\beta - \gamma_1)k^{\omega_1} + \tfrac{1}{2}[\gamma_1(\gamma_1 - 1) - \beta(\beta - 1)]k^{2\omega_1} + o(k^{2\omega_1}) \\
c_2(\Theta) &= (\gamma_2 - \beta)k^{\omega_0} + \tfrac{1}{2}[\gamma_2(\gamma_2 - 1) - \beta(\beta - 1)]k^{2\omega_0} + o(k^{2\omega_0}) \\
c_2(\theta) &= (\beta - \gamma_2)k^{\omega_1} + \tfrac{1}{2}[\gamma_2(\gamma_2 - 1) - \beta(\beta - 1)]k^{2\omega_1} + o(k^{2\omega_1}).
\end{aligned} \tag{A14}$$

The constants are

$$\begin{aligned}
C_1 &= \frac{(\beta - \gamma_2)k^{\omega_1 + \alpha_G \omega_0} - \lambda(\beta - \gamma_2)k^{\omega_0 + \alpha_L \omega_1} + o(k^{\min(\omega_0 + \alpha_L \omega_1, \omega_1 + \alpha_G \omega_0)})}{C(k^{2\omega_0 + \omega_1} + k^{2\omega_1 + \omega_0}) + o(k^{\min(2\omega_0 + \omega_1, 2\omega_1 + \omega_0)})} \\
C_2 &= \frac{(\gamma_1 - \beta)k^{\omega_1 + \alpha_G \omega_0} - \lambda(\gamma_1 - \beta)k^{\omega_0 + \alpha_L \omega_1} + o(k^{\min(\omega_0 + \alpha_L \omega_1, \omega_1 + \alpha_G \omega_0)})}{C(k^{2\omega_0 + \omega_1} + k^{2\omega_1 + \omega_0}) + o(k^{\min(2\omega_0 + \omega_1, 2\omega_1 + \omega_0)})}
\end{aligned} \tag{A15}$$

where $\quad C = \tfrac{1}{2}(\gamma_1 - \gamma_2)(\gamma_1 - \beta)(\beta - \gamma_2) \geq 0.$

If we choose the convergence rates, $\omega_0$ and $\omega_1$, such that $\omega_0/\omega_1 > (1-\alpha_L)/(1-\alpha_G)$, then from (A15) it is easy to verify that $C_1, C_2 \to \infty,$ as $k \to 0.$ So this strategy gives infinite utility in the limit and the truly optimal strategy must as well.

## D: Proof of Proposition 1

As illustrated in Figure 2 there can be two local maxima to our optimization problem.

---

[47] Unbounded utility can be achieved for scaled-TK utility by realizing a series of infinitesimally sized gains because $u'(0^+) = \infty$, and with no transactions costs, nothing prevents the investor from doing so. Unbounded utility cannot be achieved in the same way with modified-TK as its marginal utility is bounded.

[48] Given that $\omega_0$ and $\omega_1$ are both less than ½, all the terms associated with $c$ are included in the higher-order terms $o(k^{2\omega_0})$ and $o(k^{2\omega_1})$.

The one-point maximum is a corner solution with θ = 0; the two-point maximum is an interior maximum with θ > 0. Which local maxima is the global maximum depends on the specific economic and utility parameters, but it can be most easily characterized by the loss aversion parameter, λ. The value function for the one-point maximum does not depend on λ since no losses are ever realized. The value function for the two-point maximum is obviously decreasing in λ. Therefore, there is a single critical value of λ at which the two value functions are equal marking the change in regime.

Denote the two-point and one-point value functions as $v^{(2)}(x) = C_1^{(2)} x^{\gamma_1} + C_2^{(2)} x^{\gamma_2}$ and $v^{(1)}(x) = C_1^{(1)} x^{\gamma_1}$. Since, for the critical value of $\lambda_*$, the value functions are equal everywhere that they are defined, we must have $C_1^{(2)} = C_1^{(1)} = C_1$ and $C_2^{(2)} = 0$. From (9), the boundary conditions are

$$C_1 \Theta^{\gamma_1} = u(\kappa\Theta - 1) + (K\Theta)^\beta C_1 \qquad C_1 \theta^{\gamma_1} = u(\kappa\theta - 1) + (K\theta)^\beta C_1 \tag{A16}$$

The smooth-pasting conditions for a maximum are

$$\gamma_1 C_1 \Theta^{\gamma_1 - 1} = \kappa u'(\kappa\Theta - 1) + \beta K^\beta \Theta^{\beta-1} C_1 \qquad \gamma_1 C_1 \theta^{\gamma_1 - 1} = \kappa u'(\kappa\theta - 1) + \beta K^\beta \theta^{\beta-1} C_1. \tag{A17}$$

The two boundary conditions in (A16) must yield the same value for $C_1$; therefore,

$$C_1 = \frac{u(\kappa\Theta - 1)}{\Theta^{\gamma_1} - (K\Theta)^\beta} = \frac{u(\kappa\theta - 1)}{\theta^{\gamma_1} - (K\theta)^\beta}. \tag{A18}$$

Similarly, from (A17)

$$C_1 = \frac{\kappa u'(\kappa\Theta - 1)}{\gamma_1 \Theta^{\gamma_1 - 1} - \beta K^\beta \Theta^{\beta-1}} = \frac{\kappa u'(\kappa\theta - 1)}{\gamma_1 \theta^{\gamma_1 - 1} - \beta K^\beta \theta^{\beta-1}} \tag{A19}$$

Substituting the scaled-TK utility function into (A18) and solving for $\lambda_*$ gives

$$\lambda_* = -\frac{(\kappa\Theta - 1)^{\alpha_G}}{(1 - \kappa\theta)^{\alpha_L}} \frac{\theta^{\gamma_1} - K^\beta \theta^\beta}{\Theta^{\gamma_1} - K^\beta \Theta^\beta}. \tag{A20}$$

Combining (A18) and (A19) gives

$$\begin{aligned}
0 &= (\alpha_G - \gamma_1)\kappa\Theta^{\gamma_1 + 1 - \beta} + \gamma_1 \Theta^{\gamma_1 - \beta} - (\alpha_G - \beta)K^\beta \kappa\Theta - \beta K^\beta \\
0 &= (\alpha_L - \gamma_1)\kappa\theta^{\gamma_1 + 1 - \beta} + \gamma_1 \theta^{\gamma_1 - \beta} - (\alpha_L - \beta)K^\beta \kappa\theta - \beta K^\beta.
\end{aligned} \tag{A21}$$

These equations can be re-expressed as

$$\Theta^{\gamma_1} - (K\Theta)^\beta = \frac{(K\Theta)^\beta (\gamma_1 - \beta)(\kappa\Theta - 1)}{(\alpha_G - \gamma_1)\kappa\Theta + \gamma_1} \qquad \theta^{\gamma_1} - (K\theta)^\beta = \frac{(K\theta)^\beta (\gamma_1 - \beta)(\kappa\theta - 1)}{(\alpha_L - \gamma_1)\kappa\theta + \gamma_1}. \tag{A22}$$

Substituting back into (A20) gives

$$\lambda_* = \frac{(\kappa\Theta-1)^{\alpha_G-1}\theta^\beta}{(1-\kappa\theta)^{\alpha_L-1}\Theta^\beta} \times \frac{(\alpha_G-\gamma_1)\kappa\Theta+\gamma_1}{(\alpha_L-\gamma_1)\kappa\theta+\gamma_1} \tag{A23}$$

which is the desired expression in (12).

Correspondingly for modified-TK utility introduced in the calibration session, (A18) now gives

$$\lambda_* = \frac{\alpha_L}{\alpha_G}\frac{(\kappa\Theta)^{\alpha_G}-1}{(\kappa\theta)^{\alpha_L}-1}\frac{\theta^{\gamma_1}-K^\beta\theta^\beta}{\Theta^{\gamma_1}-K^\beta\Theta^\beta}, \tag{A24}$$

instead of (A20), and (A19) together with (A18) now give

$$\begin{aligned}0 &= (\alpha-\gamma_1)\kappa^{\alpha_G}\Theta^{\gamma_1+\alpha_G-\beta}+\gamma_1\Theta^{\gamma_1-\beta}-(\alpha_G-\beta)K^\beta(\kappa\Theta)^{\alpha_G}-\beta K^\beta \\ 0 &= (\alpha_L-\gamma_1)\kappa^{\alpha_L}\theta^{\gamma_1+\alpha_L-\beta}+\gamma_1\theta^{\gamma_1-\beta}-(\alpha_L-\beta)K^\beta(\kappa\theta)^{\alpha_L}-\beta K^\beta.\end{aligned} \tag{A25}$$

Together (A24) and (A25) give

$$\lambda_* = \frac{\alpha_L}{\alpha_G}\left(\frac{\theta}{\Theta}\right)^\beta\frac{(\alpha_G-\gamma_1)\kappa^{\alpha_G}\Theta^{\alpha_G}+\gamma_1}{(\alpha_L-\gamma_1)\kappa^{\alpha_L}\theta^{\alpha_L}+\gamma_1} \tag{A26}$$

in place of (A23).

As $\beta \to \gamma_1$ from below, both (A21) and (A25) yield $\theta \approx K^{\beta/(\gamma_1-\beta)} \to 0$. Substituting this into either (A23) or (A26) gives $\lambda_* \to 0$, which implies that voluntary loss realization will never take place in this limiting case since $\lambda \geq 1 > \lambda_*$. ∎

**E: Proofs of Propositions 2 to 6**

**Proof of Proposition 2:** The ultimate resolution probabilities for a single episode, which were given in (13), can be determined from the time-independent backward equation

$$0 = \tfrac{1}{2}\sigma^2 x^2 q'' + \mu x q'. \tag{A27}$$

The time-invariant probability of any event measurable in terms of the current variable $x$ is a solution to this equation. Solving (A27) with boundary conditions $q(\theta) = 0$ and $q(\Theta) = 1$ gives the probability that $x$ will reach $\Theta$ before it reaches $\theta$ conditional on the current value of $x$

$$q(x) = \frac{x^\eta - \theta^\eta}{\Theta^\eta - \theta^\eta} \qquad \text{where} \qquad \eta \equiv 1 - \frac{2\mu}{\sigma^2}. \tag{A28}$$

The ultimate resolution probabilities in (13) are $Q_G = q(1)$, $Q_L = 1 - Q_G$.

To determine the expected duration of an investment episode, note that the process $\ln x_t - (\mu - \tfrac{1}{2}\sigma^2)t$ is a martingale starting at 0. Since the times of sales are stopping times for the process, the martingale stopping time theorem gives

$$0 = \mathbb{E}[\ell n\, \tilde{x}_\tau - (\mu - \tfrac{1}{2}\sigma^2)\tilde{\tau}] = Q_G \ell n\, \Theta + Q_L \ell n\, \theta - (\mu - \tfrac{1}{2}\sigma^2)\mathbb{E}[\tilde{\tau}] \quad (A29)$$

from which (15) is immediate.

Over repeated investment episodes, the stochastic process for $x \equiv X/R$ is a Markov process with bounded support, $x \in (\theta, \Theta)$. It is a diffusion everywhere except at $x = 1$, $\theta$, and $\Theta$. It is not a diffusion at those points because whenever $x$ reaches either boundary, it returns immediately to $x = 1$ due to the sale and reinvestment. The steady-state distribution of $x$ is the solution to the Kolmogorov forward equation

$$0 = \frac{1}{2}\frac{d^2}{dx^2}[\sigma^2 x^2 f] - \frac{d}{dx}[\mu x f] = \tfrac{1}{2}\sigma^2 x^2 f_{xx} + (2\sigma^2 - \mu)x f_x + (\sigma^2 - \mu)f \quad (A30)$$

$$\text{for } x \in (\theta,1) \cup (1,\Theta) \quad \text{and subject to} \quad f(\theta) = f(\Theta) = 0.$$

The density must vanish at the boundaries just as it does with an absorbing barrier because the infinite variation in the diffusion process assures that the barrier will be reached with probability approaching unity whenever the diffusion enters a small neighborhood of the barrier.

Since $x$ is not a diffusion at 1, the differential equation does not hold at that point, and the equation must be solved separately in the two regions then pieced together. The general solution to the differential equation is $Ax^{-1} + Bx^{-1-\eta}$ where $\eta \equiv 1 - 2\mu/\sigma^2$, and the two constants differ in the two regions. The two boundary conditions, the continuity of the density at $x = 1$, and the unit mass of $f$ over the entire region supply the four equations needed to determine the four constants. The density and cumulative distribution function are

$$f(x) = \begin{cases} \dfrac{(\Theta^\eta - 1)(\theta^\eta x^{-\eta} - 1)x^{-1}}{(1-\theta^\eta)\ell n\,\Theta + (\Theta^\eta - 1)\ell n\,\theta} \\[1em] \dfrac{(\theta^\eta - 1)(\Theta^\eta x^{-\eta} - 1)x^{-1}}{(1-\theta^\eta)\ell n\,\Theta + (\Theta^\eta - 1)\ell n\,\theta} \end{cases} \text{ and } F(x) = \begin{cases} \dfrac{(\Theta^\eta - 1)\left(\ell n(\theta/x) + \tfrac{1}{\eta}[1 - (\theta/x)^\eta]\right)}{(1-\theta^\eta)\ell n\,\Theta + (\Theta^\eta - 1)\ell n\,\theta} & x \leq 1 \\[1em] 1 - \dfrac{(\theta^\eta - 1)\left(\ell n(x/\Theta) + \tfrac{1}{\eta}[(\Theta/x)^\eta - 1]\right)}{(1-\theta^\eta)\ell n\,\Theta + (\Theta^\eta - 1)\ell n\,\theta} & x \geq 1. \end{cases} \quad (A31)$$

The probabilities in equation (14) are $\varphi_G = 1 - \varphi_L = 1 - F(1)$. ∎

**Proof of Proposition 3:** It is well-known that the duration of each episode of a Poisson process has an exponential distribution with probability density $\rho e^{-\rho t}$. At the end of an episode lasting $\tau$, $\ell n\, x$ has a normal distribution with mean $(\mu - \sigma^2/2)\tau$ and variance $\sigma^2 \tau$. Therefore, the expected price ratios conditional on a sale at a gain and a loss are

$$\bar{\Theta} = (Q_G)^{-1} \int_0^\infty \rho e^{-\rho\tau} \int_0^\infty \tfrac{1}{\sigma\sqrt{\tau}} e^z \phi\!\left(\tfrac{z-(\mu-\sigma^2/2)\tau}{\sigma\sqrt{\tau}}\right) dz\, d\tau = -\frac{\rho(1-\psi^-)}{(\rho-\mu)\psi^-}$$

$$\bar{\theta} = (Q_L)^{-1} \int_0^\infty \rho e^{-\rho\tau} \int_{-\infty}^0 \tfrac{1}{\sigma\sqrt{\tau}} e^z \phi\!\left(\tfrac{z-(\mu-\sigma^2/2)\tau}{\sigma\sqrt{\tau}}\right) dz\, d\tau = \frac{\rho(\psi^+ - 1)}{(\rho-\mu)\psi^+}$$

(A32)

with $\rho > \mu$ required for $\overline{\Theta}$ to be finite.

The probability, $H(x)$, that a given episode eventually events with a loss conditional on the current asset value ratio, $x = X/R$, satisfies the modified backward equation

$$0 = \tfrac{1}{2}\sigma^2 x^2 H'' + \mu x H' + \rho(\mathbf{1}_{x<1} - H) \qquad \text{subject to } H(0) = 1, H(\infty) = 0 \qquad (A33)$$

where $\mathbf{1}_{x<1}$ is the indicator function that the asset currently has a paper loss. This final term is the probability per unit time that $H$ changes discontinuously to one (zero) if a sales event occurs when $x$ is less (greater) than 1 resulting in a realized loss (gain). Solving this equation in the two regions and matching it and its derivative at the boundary $x = 1$ gives

$$H(x) = \begin{cases} \dfrac{\psi^+}{\psi^+ - \psi^-} x^{\psi^-} & x > 1 \\ 1 + \dfrac{\psi^-}{\psi^+ - \psi^-} x^{\psi^+} & x < 1. \end{cases} \qquad (A34)$$

The probabilities that a new investment episode ends in a sale at a loss and gain are $Q_L = H(1)$ and $Q_G = 1 - H(1)$. Since the Poisson events determining sales are independent of the stochastic process of the stock price, the probability of realizing a loss for any episode must equal the steady-state probability of holding an unrealized paper loss; that is, $\varphi_L = Q_L$. Similarly for gains, $\varphi_G = Q_G = 1 - H(1)$. ∎

**Proof of Proposition 4:** In a given sample of $T$ observed sales, the observed proportion of gains realized is[49]

$$PGR = \frac{\sum_{t=1}^{T} \tilde{s}_t^G}{\sum_{t=1}^{T}(\tilde{s}_t^G + \sum_{i=1, i \neq i_t}^{\tilde{n}_t} \tilde{u}_{ti}^G)}$$

$$\tilde{s}_t^G = \begin{cases} 1 & \text{if the } t^{th} \text{ trade is a realized gain} \\ 0 & \text{if the } t^{th} \text{ trade is a realized loss} \end{cases}$$

$$\tilde{u}_{ti}^G = \begin{cases} 1 & \text{if the } i^{th} \text{ stock held in the account executing the } t^{th} \text{ trade has a paper gain} \\ 0 & \text{if the } i^{th} \text{ stock held in the account executing the } t^{th} \text{ trade has a paper loss} \end{cases} \qquad (A35)$$

$\tilde{n}_t = $ the number of stocks held in the account making the $t^{th}$ trade

and $i_t$ is the index of the stock sold in the $t^{th}$ trade.

Since each stock has an equal probability of being sold in the steady state, by the Weak Law of Large Numbers, the fraction of trades made by accounts holding $n$ shares converges to $n\pi_n / \overline{n}$

---

[49] Consistent with the empirical implementation of Odean (1998), the calculation in (A35) counts paper gains and losses only when at least one stock is sold in the account. Also (A35) assumes that each trade involves the sale of a single asset in a given account. This is justified in a continuous-time model since two less than perfectly correlated assets will never reach their sales thresholds simultaneously with probability one.

where $\pi_n$ is the fraction of investors who have $n$ stocks in their accounts, and

$$\text{plim}\, T^{-1} \sum_{t=1}^{T} \tilde{s}_t^G = Q_G$$
$$\text{plim}\, T^{-1} \sum_{t=1}^{T} \sum_{i=1, i \neq i_t}^{\tilde{n}_t} \tilde{u}_{ti}^G = \sum_n (\pi_n n/\overline{n})(n-1)\varphi_G = (\overline{n} + \sigma_n^2/\overline{n} - 1)\varphi_G \,. \quad (A36)$$

From Slutsky's Theorem, that the probability limit of a function of random variables is equal to the same function of the separate probability limits, the plim of *PGR* is the ratio of these two quantities. Similar reasoning derives the plim of *PLR*. Equation (18) follows from an additional application of Slutsky's Theorem to the ratio, *PGR/PLR*. ∎

**Proof of Proposition 5:** Because asset returns are independent and identically distributed, a sequence of investment episodes for any investor is a renewal process. Therefore, by the Elementary Renewal Theorem the average number of investment episodes per unit time in a single sequence of trades for a type $i$ investor is $1/\mathbb{E}[\tau_i]$. Since type $i$ investors hold the fraction $\pi_i n_i$ of the stocks, their trades form the fraction $\pi_i n_i/\mathbb{E}[\tau_i]$ of all trades on average, and this weight is used to compute the average statistics. In determining $\overline{\varphi}_G$, this weight is applied to the sum of the probabilities that each of the $n_i - 1$ other stocks not traded by the given type $i$ investor has a paper gain. The probability limit then follows from Slutsky's Theorem. The same reasoning applies to *PGR* and *PLR*. ∎

**Proof of Proposition 6:** This proof is similar to that for Proposition 5. Here the stocks in group $i$ form the fraction $n_i/\mathbb{E}[\tau_i]$ of all trades on average, and this weight is used to compute the average statistics. In determining $\overline{\varphi}_G$, this weight is applied to the sum of probabilities that each of the $N - 1$ stocks not traded has a paper gain when one of the $n_i$ stocks is traded. The probability limit then follows from Slutsky's Theorem. The same reasoning applies to *PGR* and *PLR*. ∎

**F: Proof of Proposition 7**

To prove this proposition, we use the following lemma.

**Lemma:** The last two conditions (iii) and (iv) in Proposition 7

$$\text{(i)}\ \delta > 0 \quad \text{(ii)}\ \frac{\partial U}{\partial G} > 0 \quad \text{(iii)}\ \frac{\partial^2 U}{\partial G^2} \leq 0 \quad \text{(iv)}\ \frac{\partial^2 U}{\partial G\, \partial R} \geq 0 \quad (A37)$$

lead to the two relations

$$U(X-R,R) \leq \int_R^X \frac{\partial U(0^-, r)}{\partial G}\, dr \quad \text{if } X < R, \quad U(X-R,R) \leq \int_R^X \frac{\partial U(0^+, r)}{\partial G}\, dr \quad \text{if } X > R. \quad (A38)$$

**Proof:** For $X > R$, we have

$$U(X-R,R) = \int_R^X \frac{\partial U(r-R,R)}{\partial G}\, dr \leq \int_R^X \frac{\partial U(0^+, R)}{\partial G}\, dr \leq \int_R^X \frac{\partial U(0^+, r)}{\partial G}\, dr \,. \quad (A39)$$

The first inequality holds because marginal utility is decreasing (assumption iii), and the second integrand is evaluated at $0^+$, its smallest argument in the range of the integral. The second inequality holds because marginal utility is increasing in $R$ (assumption iv) and each of the reference levels at which the integrand is evaluated is above $R$. A similar proof holds when $X < R$. ∎

**Proof of Proposition 7:** The reference level for the $k^{\text{th}}$ sale is $R_k = X_{k-1}$ since there are no transaction costs, the reference level is constant between sales, and the reference level is updated to the investment value after a sale. For any given realized stochastic path of $X$, assume there is a realized loss, and denote its reference level as $R_I = X_{I-1}$. By assumption, $X_I < X_{I-1}$. First, assume that the asset is eventually sold at a price above the reference level for this loss. Let $J$ denote the first such sale; that is, for some $J$, $X_J > X_{I-1} > X_{J-1}$. The total utility realized from the sales $I$ through $J$ is

$$\sum_{k=I}^{J} e^{-\delta t_k} U(X_k - X_{k-1}, X_{k-1}) = \sum_{k=I}^{J-1} e^{-\delta t_k} U(X_k - X_{k-1}, X_{k-1}) + e^{-\delta t_J} U(X_{I-1} - X_{J-1}, X_{J-1}) \\ + e^{-\delta t_J} U(X_J - X_{J-1}, X_{J-1}) - e^{-\delta t_J} U(X_{I-1} - X_{J-1}, X_{J-1}). \quad (A40)$$

The equality is a tautology with the utility from a fictitious sale occurring at time $t_J$ at value $X_{I-1}$ both added and subtracted. This lets us split the sales into two parts. The first line includes sales for which the gains and losses (though not their utility) net exactly to zero. The utility of this part is bounded by applying the lemma

$$\sum_{k=I}^{J-1} e^{-\delta t_k} U(X_k - X_{k-1}, X_{k-1}) + e^{-\delta t_J} U(X_{I-1} - X_{J-1}, X_{J-1}) \\ \leq \sum_{\substack{k=I \\ X_k < X_{k-1}}}^{J-1} e^{-\delta t_k} \int_{X_{k-1}}^{X_k} \frac{\partial U(0^-, x)}{\partial G} dx + \sum_{\substack{k=I \\ X_k > X_{k-1}}}^{J-1} e^{-\delta t_k} \int_{X_{k-1}}^{X_k} \frac{\partial U(0^+, x)}{\partial G} dx \\ + e^{-\delta t_J} \int_{X_{J-1}}^{X_{I-1}} \frac{\partial U(0^+, x)}{\partial G} dx < 0. \quad (A41)$$

This contribution to utility is negative since each element of the first integrand over losses has a matching element in the second or third integrand of gains. Note that the integrals in the first sum are negative since the lower limits are larger than the upper limits and the integrand is strictly positive by assumption (ii). Assumptions (i) through (iv) assure that the marginal utility of each loss element is greater than that of the matching gain element. In particular, marginal utility is decreasing in $G$ (assumption iii) so evaluating the marginal utilities of gains at $0^+$ makes them smaller than the marginal utilities of losses evaluated at $0^-$. Marginal utility is also decreasing in time (assumption i and ii) and increasing in $R$ (assumption iv), and each gain occurs later and has a smaller reference level than the corresponding loss.[50]

Since the terms in the first line of (A40) are negative, the total utility realized from the

---

[50] The final inequality is strict because condition (i) on δ and condition (ii) on the marginal utility are strict.

sales *I* through *J* is less than the sum of the two terms in the second line

$$\sum_{k=I}^{J} e^{-\delta t_k} U(X_k - X_{k-1}, X_{k-1}) < e^{-\delta t_J} U(X_J - X_{J-1}, X_{J-1}) - e^{-\delta t_J} U(X_{I-1} - X_{J-1}, X_{J-1})$$

$$= e^{-\delta t_J} \int_{X_{I-1}}^{X_J} \frac{\partial U(x - X_{J-1}, X_{J-1})}{\partial G} dx \leq e^{-\delta t_J} \int_{X_{I-1}}^{X_J} \frac{\partial U(x - X_{I-1}, X_{I-1})}{\partial G} dx \quad \text{(A42)}$$

$$= e^{-\delta t_J} U(X_J - X_{I-1}, X_{I-1}).$$

The new inequality in the second line holds because of assumptions (iii) and (iv). The final comparison shows that holding the stock with reference level $X_{I-1}$ and selling at $X_J$ lead to strictly higher utility than that provided by the original plan, which, therefore, cannot be optimal.

Now suppose there is no later sale that occurs at a level above $X_{I-1}$. In this case, the accumulated utility up through every sale *J* after *I* is weakly negative. Precisely, using the same reasoning as before,

$$\sum_{k=I}^{J} e^{-\delta t_k} U(X_k - X_{k-1}, X_{k-1}) < \sum_{\substack{k=I \\ X_k < X_{k-1}}}^{J} e^{-\delta t_k} \int_{X_{k-1}}^{X_k} \frac{\partial U(0^-, x)}{\partial G} dx + \sum_{\substack{k=I \\ X_k > X_{k-1}}}^{J} e^{-\delta t_k} \int_{X_{k-1}}^{X_k} \frac{\partial U(0^+, x)}{\partial G} dx < 0. \text{ (A43)}$$

This is true even for an infinite investment horizon with $J \to \infty$. Therefore, holding the stock at a reference level of $X_{I-1}$ with no subsequent sales is strictly better than the original plan. ∎